\documentclass[subscriptcorrection,upint,varvw,barcolor=Goldenrod3,mathalfa=cal=euler,balance,hyphenate,french, nolists, nofoot]{arxiv_template} %

\usepackage{algpseudocode}
\usepackage{outlines}
\usepackage{nicematrix,tikz}
\usepackage{arydshln}
\usepackage{float}
\usepackage{lipsum}
\usepackage{capt-of}
\usepackage{caption}
\usepackage{expl3}
\DeclareMathAlphabet{\mathcal}{OMS}{cmsy}{m}{n}
\usepackage{booktabs} 
\usepackage{makecell}
\usepackage{algorithm2e}
\usepackage{changepage}
\usepackage[noblocks]{authblk}  
\usepackage[absolute,overlay]{textpos} 

\hypersetup{%
	pdfauthor={Nomi Yu, Ferdous MD Alam, A John Hart, Faez Ahmed},                       		   	
	pdftitle={GenCAD-3D},                  	
	pdfkeywords={Generative Design, CAD, Design Automation, Imbalanced Data},
	pdfsubject = {GenCAD-3D},			
}

\JourName{Mechanical Design}
                   
\begin{document}

\begin{textblock*}{20cm}(1cm,1.5cm) 
    \centering
    \large \textbf{\color{gray} This is a preprint version of the manuscript. \\ The final published version is available in the Journal of Mechanical Design via https://doi.org/10.1115/1.4069276} \\
\end{textblock*}

\author[1]{\Large Nomi Yu}
\author[1]{\Large Md Ferdous Alam}
\author[1]{\Large A John Hart} 
\author[1]{\Large Faez Ahmed} 
\date{}
\affil[1]{\large Massachusetts Institute of Technology}

\title{
    \rule{17cm}{1.5pt}\\
    \textbf{\fontsize{18}{18}\selectfont{GenCAD-3D:} \\
    \fontsize{16}{16}\selectfont{CAD Program Generation using  \\Multimodal Latent Space Alignment and \\Synthetic Dataset Balancing} \\
    \rule{17cm}{1.5pt}
}}

\twocolumn[
\begin{@twocolumnfalse}
\maketitle 

\begin{adjustwidth}{10mm}{10mm}

\begin{abstract}
\normalsize

CAD programs, structured as parametric sequences of commands that compile into precise 3D geometries, are fundamental to accurate and efficient engineering design processes. Generating these programs from nonparametric data such as point clouds and meshes remains a crucial yet challenging task, typically requiring extensive manual intervention. Current deep generative models aimed at automating CAD generation are significantly limited by imbalanced and insufficiently large datasets, particularly those lacking representation for complex CAD programs. To address this, we introduce GenCAD-3D, a multimodal generative framework utilizing contrastive learning for aligning latent embeddings between CAD and geometric encoders, combined with latent diffusion models for CAD sequence generation and retrieval. Additionally, we present SynthBal, a synthetic data augmentation strategy specifically designed to balance and expand datasets, notably enhancing representation of complex CAD geometries. Our experiments show that SynthBal significantly boosts reconstruction accuracy, reduces the generation of invalid CAD models, and markedly improves performance on high-complexity geometries, surpassing existing benchmarks. These advancements hold substantial implications for streamlining reverse engineering and enhancing automation in engineering design. We will publicly release our datasets and code, including a set of 51 3D-printed and laser-scanned parts on our project site\footnotemark{}.

\end{abstract}
\end{adjustwidth}

\vspace{0.5cm}

\end{@twocolumnfalse}
]

\captionsetup{font=sf}


\begin{figure*}[!h]
    \centering
    \includegraphics[width=0.95\linewidth]{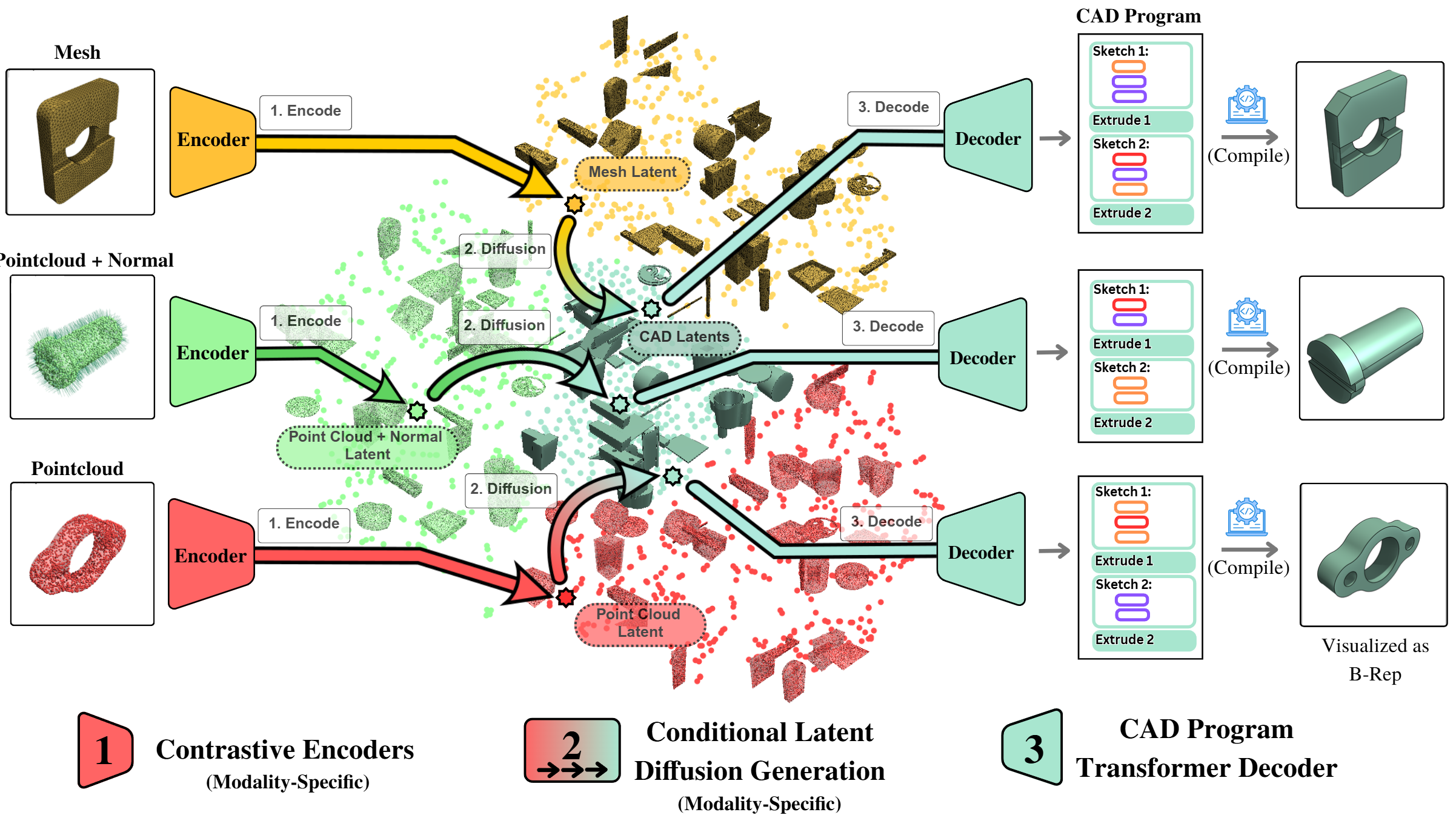}
    \caption{\textbf{Multimodal Generation through Latent Space Alignment.} We visualize the joint latent space of multiple modalities through t-SNE. The output of our model is a CAD program, which is visualized by its compiled B-Rep.}
    \label{fig:teaser}
\end{figure*}

\section{Introduction}
Computer-Aided Design (CAD) is integral to engineering design and manufacturing, underpinning the creation and modification of nearly all modern products, from automobiles and consumer electronics to everyday household items~\citep{weiler1986topological, mantyla1987introduction}.

\begin{figure}
    \centering
    \vspace{-10mm}
    \includegraphics[width=1\linewidth]{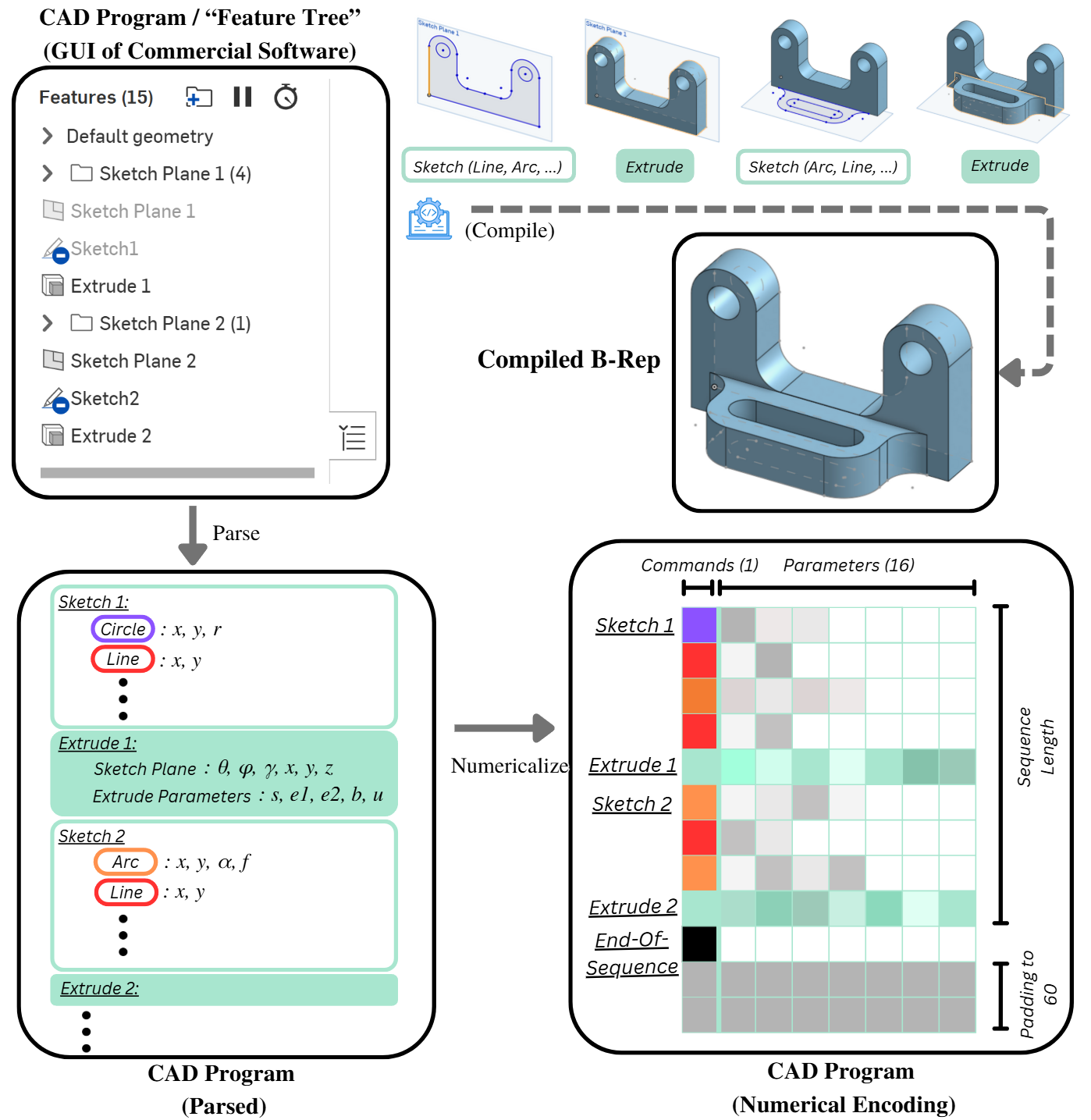}
    \caption{\textbf{A CAD program consists of a sequence of commands and parameters used to compile a 3D B-Rep object.} CAD programs often use a sketch-and-extrude strategy: first a 2D sketch is drawn using Line, Arc, and Circle commands, which is then extruded into a 3D solid via an Extrude command. We encode the CAD program into a $60\times17$ matrix for compatibility with our autoencoder.}
    \label{fig:cad-program}
\end{figure}

A CAD model (see Fig. \ref{fig:cad-program}) is defined by a sequence of parametric commands commonly known as a CAD program or "Feature Tree," which can then be compiled into an exact 3D solid model known as a Boundary Representation model (B-Rep). While the final output is the B-Rep model, it is the CAD program representation that provides engineers with the precise, parametric control over geometric manipulation that is not offered by other 3D modalities. CAD programs are integral to engineering design due to their editability and integration with other steps of the design process. For example, industry engineers often pursue reverse engineering to convert 3D scans (point clouds) of physical objects into editable CAD programs. This is critical for modifying and manufacturing parts when a CAD program is unavailable. For example, when a part in a decades-old machine breaks and needs to be replaced, but the original supplier has discontinued that part, an engineer would laser-scan the broken part, reverse-engineer that scan into a CAD program, then use that model for downstream manufacturing. While current software tools can assist this reverse-engineering process by detecting and fitting analytical surfaces to features in the scan \cite{Buonamici04052018}, these tools still demand substantial expert input and decision-making, making the process expensive and time‑consuming.

\footnotetext{\url{gencad3d.github.io}}

Meanwhile, generative AI techniques have made strides in producing 3D geometries, but they typically output shapes in non-parametric forms like meshes or implicit surfaces. Such free-form outputs, while often visually close to a user’s request, lack the parametric structure of CAD and are thus difficult to edit or refine precisely. In practice, a generated 3D model might require manual adjustments to meet exact design intent, which is especially problematic for engineering applications that demand strict tolerance---for instance, to achieve precise fits or alignments in mechanical assemblies \cite{wei2022chain, liu2024alignment}. Converting generative outputs into CAD programs would address these issues by reintroducing parametric control. Unfortunately, manually creating CAD programs is itself a complex skill that typically takes years to master~\citep{piegl2005ten}. Even experienced designers spend considerable time on repetitive, boilerplate modeling steps. An automated system for generating CAD programs from geometry would thus be revolutionary in lowering the barrier to CAD modeling and reducing manual workload \citep{thompson1999feature, beniere2013comprehensive, varady1997reverse}.

Deep generative models offer a promising avenue for automating CAD program creation. These models have already excelled at cross-modal tasks (such as text-to-image generation) by learning shared latent representations across modalities. In the CAD domain, early research has explored generating simple CAD programs. For example, DeepCAD introduced a deep network for unconditional CAD program generation \cite{wu2021deepcad}, and subsequent work demonstrated CAD program generation conditioned on 2D images or sketches \cite{alam2024gencad}. However, there has been limited progress in generating CAD programs directly from 3D geometric data (point clouds or meshes) \cite{xu2024cad}. A primary obstacle is the scarcity of training data for complex CAD programs. One of the largest public datasets, DeepCAD, provides many CAD programs but is heavily skewed toward low-complexity models. High-complexity CAD programs (with many modeling steps) are underrepresented, causing models trained on this data to excel on simple shapes but struggle with more intricate ones. In fact, standard evaluation metrics, which average performance over the whole dataset, are dominated by results on simple examples. This means improvements (or failures) on complex geometries can be masked, hindering the development of models that genuinely perform well on high-complexity designs.

In this work, we introduce GenCAD-3D, a multimodal generative framework that addresses these challenges through multimodal latent space alignment and synthetic dataset balancing. Our main contributions can be summarized as follows:

\begin{itemize}

\item \textbf{GenCAD-3D Framework:} We introduce GenCAD-3D, a novel generative framework that utilizes modality-specific encoders combined with contrastive learning to align a shared latent space across CAD programs and 3D geometric representations. This aligned representation, together with a conditional latent diffusion model, enables robust cross-modal retrieval, accurate geometry-to-CAD reconstruction, and efficient generative modeling of CAD sequences
\item \textbf{Specialized 3D Encoders:} We propose and evaluate specialized neural encoders for mesh inputs, complementing the point cloud encoder. Compared to a point cloud-only baseline, these mesh-aware encoders yield up to 15\% relative improvement in command accuracy and 3\% improvement in parameter accuracy during conditional reconstruction and up to an 11\% improvement in top-1 CAD retrieval accuracy, with the gains most pronounced at large retrieval scales (searching among 2048 CAD models). This result highlights the benefit of exploiting mesh information for better shape understanding in the latent space.

\item \textbf{Synthetic Dataset Balancing and Expansion (SynthBal)}: We introduce SynthBal, a synthetic data augmentation strategy designed to correct the imbalance in CAD program complexity within the training dataset and to artificially increase dataset size. By balancing and expanding the original dataset, SynthBal greatly improves all of our models' performances on complex shapes. Notably, it reduces the rate of invalid CAD generations from 3.44\% (with the original dataset) to 0.845\%, and decreases the median Chamfer distance error by up to 89\%, especially for long (high-complexity) CAD programs. We also see performance improvements not only in the models trained with SynthBal but also in downstream tasks that utilize them. These improvements demonstrate the effectiveness of using synthetic data to compensate for the long-tail distribution of CAD data.

\item \textbf{Complexity-Normalized Evaluation:} We develop an evaluation metric that normalizes performance across different sequence lengths, providing a fair assessment of reconstructive accuracy for both simple and complex CAD programs. This sequence-length normalized metric offers a reflection of the model’s capability on high-complexity CAD generation, ensuring that progress on challenging cases is captured and encouraging models that perform well across the full complexity spectrum.

\item \textbf{Public Datasets and Methods:} We release the following to enable further research in reverse-engineering and deep learning for CAD: our multimodal geometric dataset and encoders, our SynthBal augmentation strategy, and a multimodal dataset of 51 real-world scans of 3D printed parts and their corresponding CAD programs. This scanned dataset is essential since the physical process of laser-scanning introduces artifacts that are nontrivial to virtually replicate. 
\end{itemize}


\section{Related Work}\label{sec:relatedwork}
This section reviews prior work relevant to GenCAD-3D, including reverse engineering methods for CAD reconstruction, multimodal latent space alignment strategies, generative models for 3D geometries and CAD sequences, and approaches addressing data imbalance in CAD datasets.

\subsection{Reverse Engineering}
Reverse engineering---converting geometric inputs such as scanned point clouds into editable CAD programs---is a longstanding challenge in the engineering community \cite{Buonamici04052018}. Traditionally, reverse engineering techniques use analytical methods that segment geometric features from point clouds, then fit topological surfaces to these segmented features. Software tools, including some neural network-based methods, have automated some of these steps, but most of these tools are only capable of generating B-Rep surfaces or 2D sketch contours \cite{sharma2020parsenetparametricsurfacefitting, liu2023point2cadreverseengineeringcad}. Achieving parametric CAD programs from these B-Rep outputs usually requires significant manual input and expert intervention, and so a fully automated and robust method capable of generating complete parametric CAD programs from geometric inputs remains elusive. Unlike prior work, GenCAD-3D provides a fully automated method capable of directly generating complete parametric CAD programs from geometric inputs without manual intervention. 

\subsection{Generative Models for 3D Geometries and CAD}
Generative models have significantly advanced in the context of 3D geometry, covering a variety of geometric representations such as meshes\citep{tang2023dreamgaussian, jun2023shap, liu2024one, ramesh2021zero, gan_3d, deep_gen_review}, point clouds \citep{sun2020pointgrow}, B-reps \cite{jayaraman2023solidgen, xu2024brepgen}, and constructive solid geometries \cite{sharma2018csgnetneuralshapeparser}.

Focus has recently shifted to CAD program generation due to its potential in the automation of many design tasks. Wu et al developed DeepCAD \cite{wu2021deepcad}, a transformer-based latent generative adversarial network (l-GAN) model, that can generate unconditional CAD. The following works have built upon unconditional CAD generation by addressing issues in data scarcity and error propagation, 
\cite{xu2022skexgen, jung2024contrastcad}. Other works have focused specifically on improving CAD sketch generation, such as by enforcing sketch constraints \cite{para2021sketchgen}. Across all 3D modalities, including CAD programs, research on 3D generative models has focused on transformer-based architectures due to their effectiveness in modeling sequences of operations or features \citep{xu2022skexgen, para2021sketchgen, wu2021deepcad}. 

Recent work has also explored conditional CAD program generation from text, images, and sketches \cite{you2024img2cadreverseengineering3d, 10.1145/3528223.3530133, wang2024vqcad, xu2024cad, cad_finetuned, alam2024gencad}. Research has also begun exploring generation from 3D modalities like pointclouds. CAD-MLLM \cite{xu2024cad} has demonstrated conditional generation from point clouds by aligning multimodal embeddings from foundational models. We create our own embeddings, allowing us to align modalities for which foundational models have not been developed, like meshes. Point2Cyl \cite{uy2022point2cyl}, CAD-SIGNet \cite{khan2024cad}, and TransCAD \cite{dupont2025transcad} have demonstrated CAD program reconstruction from point clouds by using neural networks to explicitly identify 2D sketch faces and their extrusion parameters from the pointcloud. While these methods show strong results, they are strictly limited to sketch-and-extrude CAD programs and cannot generalize to common industry commands like revolves and fillets. While our study also focuses sketch-and-extrude CAD programs, we highlight that our architecture does not make any assumptions on the type of commands used and can be generalized to a more diverse command library.

\subsection{Multimodal Learning for 3D Geometries}
Multimodal neural representation learning has attracted increasing attention due to its promise in cross-modal retrieval, alignment, and generation tasks for 3D shape processing. 
Recent literature demonstrates that contrastive loss is particularly effective in aligning multiple data modalities (typically text and images) in the latent space \citep{radford2021learning}. Approaches inspired by the Contrastive Vision-Language Pre-training (CLIP) method have been used to align point cloud and image modalities \citep{zhang2022pointclip, ghose2025clip, 10.1115/1.4067085}. Similarly, contrastive loss has been used to align point cloud, text, and images to CAD programs using multimodal LLMs \citep{xu2024cad, dai2022enabling}. As previously mentioned, these methods depend on existing foundational models, unlike our method. Recently, wavelet latent diffusion has been used for multimodal B-rep generation \citep{sanghi2024wavelet}. 
Unlike previous methods, GenCAD-3D provides a modular architecture, that employs contrastive learning tailored specifically for mesh and point cloud modalities, independently learning multimodal embeddings without relying on existing foundational models.

\subsection{Imbalanced Data}
There have been a few works addressing imbalanced data for CAD. Most notably, the DeepCAD dataset \cite{wu2021deepcad} utilizes augmentations through a "random replacement" strategy to augment the dataset during training. More recently, ContrastCAD \cite{jung2024contrastcad} describes using a similar method with additional augmentations and a novel autoencoder training architecture. Both methods show improvements in reconstruction performance from the original by increasing diversity, but neither addresses the inherent imbalance in the dataset with respect to complex CAD programs. Although there is literature on addressing long-tailed dataset imbalance for classification models \cite{yebin2024synaugexploitingsyntheticdata}, there is a lack of work that integrates these ideas for CAD program datasets nor autoencoder reconstruction models. Addressing these limitations, our proposed SynthBal strategy explicitly tackles dataset imbalance by enhancing representation of complex CAD programs, improving reconstruction accuracy across varying levels of complexity.
In summary, there is no prior work that has combined multimodal latent alignment with a diffusion-based generator for CAD sequences, nor addressed the long-tail complexity issue – GenCAD-3D aims to fill these gaps.


\section{Representation Learning Architecture} 
\begin{figure*}
    \centering
    \includegraphics[width=0.8\linewidth]{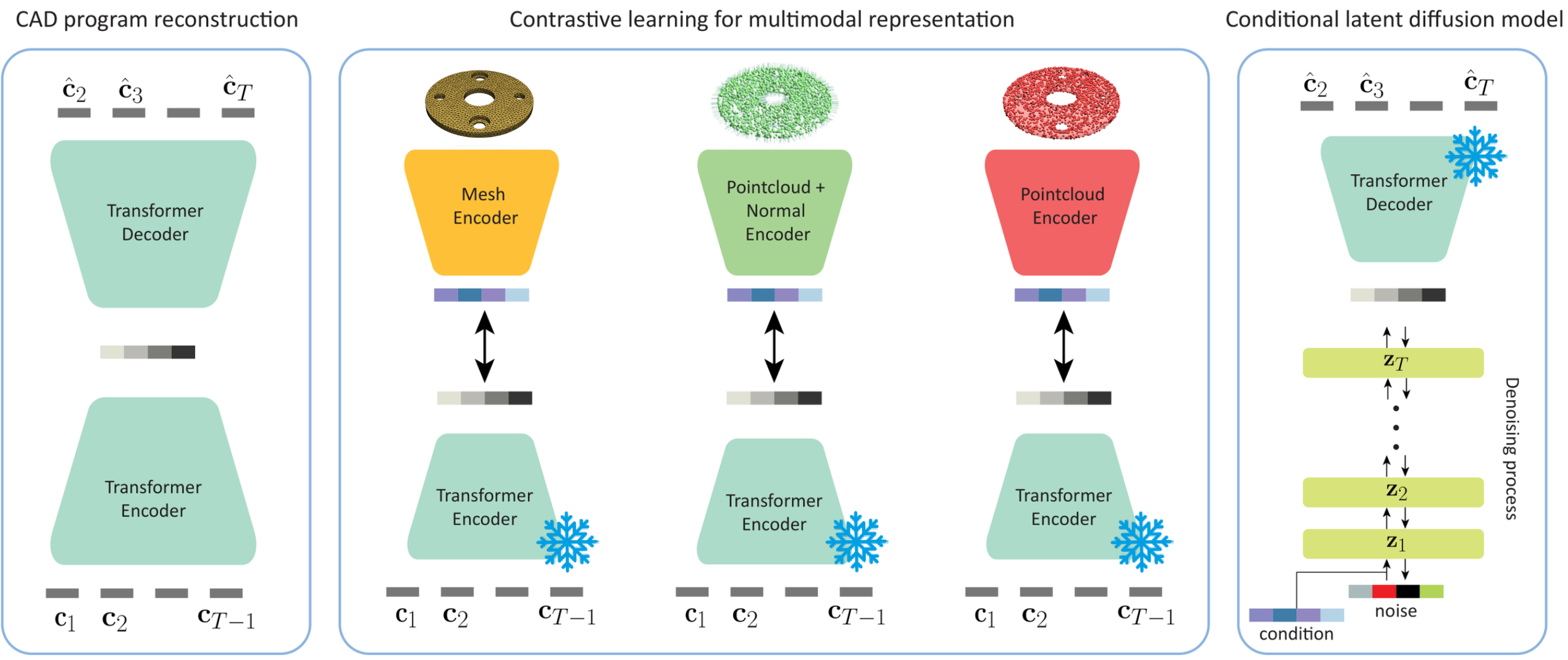}
    \caption{\textbf{Representation Learning Architecture.} The CAD program transformer autoencoder is trained once to learn a representation for CAD programs, and separate contrastive and diffusion models are trained for each of the modalities to learn their respective representations.}
    \label{fig:representation}
\end{figure*}

Our framework is inspired by the GenCAD architecture \citep{alam2024gencad}, which proposes a conditional latent diffusion model to create CAD programs from image inputs. We extend this method to 3D input modalities using contrastive multimodal representation learning and latent diffusion modeling. See Fig. \ref{fig:representation} for each component of the architecture, and Fig. \ref{fig:teaser} for the combined architecture.

\subsection{Multimodal Contrastive Learning}\label{sec:method_1}
We first align embeddings across the multiple modalities of point clouds, meshes, and CAD programs in the latent space through contrastive learning. This process contains two stages: first, learning a latent space for CAD programs through a causal transformer autoencoder  (Fig. \ref{fig:representation}-1), and second, learning latent spaces for geometric inputs by aligning them to the CAD latent space through contrastive loss (Fig. \ref{fig:representation}-2). In a properly aligned latent space, embeddings corresponding to the same geometry across modalities are pulled together while embeddings corresponding to different geometries are pushed apart. The focus of our paper is on the second stage: developing 3D modality-specific encoders to learn joint geometry-CAD program representations. 

For the first stage of CAD program representation learning, our goal is to map the CAD program into a continuous latent space $\mathbf{z}_{\mathcal{C}} \in \mathbb{R}^{d_m}$. We keep the causal transformer autoencoder architecture the same as \citep{alam2024gencad}, which has demonstrated good performance by enforcing causal relationships between commands during the decoding step. We can represent a CAD program $\mathcal{C}$ (Fig. \ref{fig:cad-program}) by a sequence of $\ell$ vectors $\mathbf{c}_i$ with fixed dimension $d$:

\begin{equation}
    \mathcal{C} = (\mathbf{c}_1, \mathbf{c}_2, \dots, \mathbf{c}_{\ell}) \; \; \; \text{where } \mathbf{c}_i \in \mathbb{R}^d, 1\leq i\leq \ell,
\end{equation}

where each vector $\mathbf{c}_i$ represents a single step, or parameterized command, of the CAD program, and $\ell$ is the \textbf{sequence length}, or number of commands, of the CAD program (more details in Sec. \ref{cad-modality}). Beginning with learned position tokens $(\mathbf{t}_1, \mathbf{t}_2, \dots, \mathbf{t}_{\ell})$, we feed the tokens through multiple transformer decoder layers while using a causal mask in which each token $\mathbf{t}_i$ can attend to preceding tokens $\mathbf{t}_0...\mathbf{t}_{i-1}$. This maintains the sequential relationship between commands in a CAD program. 

For the second stage of multimodal contrastive learning, our goal is to learn a continuous latent space for geometries $\mathbf{z}_{\mathcal{M}} \in \mathbb{R}^{d_m}$ that is aligned with the latent space for CAD programs $\mathbf{z}_{\mathcal{C}}$. Note that both these spaces have the same dimensionality $d_m$. For each modality, we utilize a modality-specific encoder, $\mathcal{E}$, that maps the input geometry, $\mathcal{M}$, into a latent space, 
\begin{equation}
    \mathbf{z}_{\mathcal{M}} = \mathcal{E}(\mathcal{M}).
\end{equation}

Unlike the CAD autoencoder, we do not learn the latent representation of the input modality using unsupervised learning; rather, it is jointly learned using contrastive loss. Given a batch of $B$ example pairs, $\{(\mathcal{C}_1, \mathcal{M}_1), \dots, (\mathcal{C}_B, \mathcal{M}_B)\}$, we have $2B$ data points in total for the contrastive prediction. By considering each corresponding pair $(\mathcal{C}_i, \mathcal{M}_i)$ as the only positive data samples, we consider the rest of the $2(B-1)$ samples as negative data pairs. If the similarity between the data pairs is expressed as cosine similarity, $\text{p}(\mathbf{u}, \mathbf{v}) = \mathbf{u}^T\mathbf{v}/||\mathbf{u}|| \mathbf{v}||$, then the contrastive loss can be defined as the following,
\begin{equation}
\ell_{i, j} = -\log \frac{\text{exp}(p(\mathbf{z}_{\mathcal{C}, i}, \mathbf{z}_{\mathcal{M}, j})/\tau)}{\sum_{k=1}^{2B} \mathbb{I}_{[k\neq i]} \text{exp}(p(\mathbf{z}_{\mathcal{C}, i}, \mathbf{z}_{\mathcal{M}, k})/\tau)},
\end{equation}

where $\tau$ is a learned temperature parameter. This contrastive loss aligns the modality-specific latent space with the CAD-program latent space. Note that we keep the encoder part of the CAD autoencoder frozen during the contrastive training, which allows the framework to scale to large-scale datasets. 

\subsection{Conditional Generation through Latent Diffusion}

The last step of our framework enables conditional CAD program generation by using diffusion to project embeddings from the previously aligned latent space of geometries into the latent space of CAD programs (Fig. \ref{fig:representation}-2). By utilizing the modality-specific latents, we train a conditional latent diffusion model that can be expressed as the following, 
\begin{equation}
    p(\mathcal{C}|\mathcal{M}, \mathcal{E}) = \underbrace{p(\mathcal{C}|\mathbf{z}_{\mathcal{C}})}_{\text{decoder}}\underbrace{p(\mathbf{z}_{\mathcal{C}}|\mathbf{z}_{\mathcal{M}})}_{\text{prior}}.
\end{equation}

This model contains two main components: a prior network to generate a CAD program latent, $\mathbf{z}_{\mathcal{C}}$, conditioned on a geometry latent, $\mathbf{z}_{\mathcal{M}}=\mathcal{E}(\mathcal{M})$; and a decoder network to convert that CAD program latent back into a CAD-program.
The prior is the neural network trained in this generative step using the previously learned geometric latent space, and the decoder is the neural network previously trained in the autoencoding step. 

To enable the generative capabilities of the prior network, we use a conditional diffusion model. The forward diffusion process, which adds noise to the latent vector in a predefined scheduler, follows directly from \cite{song2020denoising}. During the reverse diffusion process, we use a ResNet-MLP denoising model based on \cite{gorishniy2021revisiting} using multiple ResNet-MLP blocks and a normalization and linear head. To enable the conditioning, we concatenate the input geometric latent with a sampled CAD program latent and then project it to a fixed dimension as the input to the denoising model. The diffusion process can be modeled using the following loss function, 
\begin{equation}
    L = \mathbb{E}_{\mathbf{z}_{\mathcal{C}}, \mathbf{z}_{\mathcal{M}}, \epsilon \sim \mathcal{N}(0, 1), t}\left[ || \epsilon - \epsilon_\theta(\mathbf{z}_t, t, \mathbf{z}_{\mathcal{M}}) ||_2^2 \right], 
\end{equation}

where $\epsilon_\theta(\mathbf{z}_t, t, \mathbf{z}_{\mathcal{C}})$ is a time-conditional MLP with a residual connection that takes the conditional latent vector, $\mathbf{z}_{\mathcal{M}}$, as input. For each modality, we train a different prior network using the CAD program and geometry embeddings obtained from the contrastive learning step.

\section{Modalities}

\subsection{CAD Program} \label{cad-modality}
Within CAD software, a CAD model is represented as a program of commands and their associated parameters (see Fig. \ref{fig:cad-program}). While commercial software offers a myriad of command, such as fillets, lofts, revolves, for this study we limit our commands to \texttt{Line}, \texttt{Arc}, and \texttt{Circle} commands to create sketches, and an \texttt{Extrusion} command to turn sketches into 3D solids. This simplification allows us to focus on the generative reconstruction architecture of this study. Nevertheless, we highlight that this architecture allows for further extension to other more complex and commonly used commands. We also include a special \texttt{End Of Sequence} command to signal the end of the program, which we reuse for padding.

To encode this representation in a numerical format, we follow the method introduced by DeepCAD \cite{wu2021deepcad}. For a given CAD program, $\mathcal{C}$, the numerical encoding is represented by a matrix representing a sequential list of commands and their parameters, $\mathcal{C} \in \mathbb{R}^{d \times \ell}$ where $\ell$ is the sequence length of the CAD program (AKA number of commands) and $d$ is the dimension of the parameterized CAD operation. As the number of commands to represent a CAD operation varies from operation to operation, the final vector is padded to create a fixed $60\times17$ dimensional matrix for all operations. The first column corresponds to commands, while the following 16 correspond to parameters of each command. As with traditional CAD software, this numerical representation can then be reinterpreted into software-specific commands and compiled into a B-Rep object. We use OpenCascade \citep{opencascade}, an open-source geometry kernel, to perform this compilation: starting from the beginning of a sequence of commands, each command is executed one by one until the end of the program is reached.

\subsection{Point Cloud}
To encode point clouds, we use the DGCNN model \cite{wang2019dynamicgraphcnnlearning}. DGCNN is a graph neural network that applies convolutions across "dynamic graphs" constructed across neighboring points via K-nearest neighbors (KNN). We use a larger version of the original implementation of DGCNN, motivated by the fact that we are predicting an embedding rather than performing a classification. For the point cloud input, we used 3D vertex positions as features and fed them into four dynamic graph convolution layers with 128, 128, 256, 512 channels and a KNN size of 20. We then concatenate the embeddings from each intermediate layer and both max and mean-pooling to convert the vertex embeddings to a single global embedding. We finally apply MLPs of sizes 1024, 512 with dropouts of $p=0.2$. This produces an encoder model with $2.598 \times 10^6$ parameters. We trained the contrastive model for 300 epochs, with batch size 64 and learning rate 1e-3, and 2 steps of gradient accumulation on 1 H100 with 80GB VRAM. To improve robustness during training, we collected a dataset with an excess of 4096 points and bootstrapped a subsample of $2048$ points as input for each forward pass. We trained the generative model for $1\times 10^6$ epochs using a batch size of $2048$, a learning rate of $1\times 10^{-5}$, and $2$ gradient accumulation steps on one H100 GPU.

\subsection{Point Cloud + Normals (PC+N)}
Unlike point clouds, meshes provide additional information about a 3D geometry’s surface, such as surface normals. We can thus directly extend the aforementioned point cloud encoder to meshes by providing surface normals as an additional input feature. We can then encode a mesh by sampling a cloud of points and their normals from a mesh surface and concatenating them to produce a 6D input. This method is advantageous over using mesh vertices directly, as we can produce uniform surface sampling regardless of mesh discretization. We use the same DGCNN model and training method as described in the point cloud case, except for the additional surface normal inputs. This produces $2.599\times 10^6$ parameters for the encoder.

\subsection{Mesh}
Meshes also provide surface information by defining node adjacencies as edges. This allows local geometric features on a mesh surface to be preserved exactly as graphs, as opposed to being approximated by Euclidean neighborhoods. To encode meshes, we use FeaStNet \cite{verma2018feastnetfeaturesteeredgraphconvolutions}, a graph neural network utilizing "Feature-Steered" convolutions to operate on irregular and spatially defined structures. FeaStNet takes mesh inputs as collections of vertices and edges. We modify the original FeaStNet implementation, which predicted vertex embeddings for segmentation tasks. We append a mean-pooling layer at the end to reduce the vertex embeddings to a single global embedding. To account for nonuniform vertex distributions over the mesh surface, we use a weighted mean based on vertex density, as described in DiffusionNet \cite{sharp2022diffusionnetdiscretizationagnosticlearning}. For the vertex input, we use a $6D$ feature of vertex positions and vertex normals. We used FeaSt-convolutional layers of 16, 32, 64, 128, 256 channels with a library of $10$ filters (M=10). The outputs are then fed through MLP layers of sizes 512, 256, 256, using dropouts of $p=0.2$. This produced $2.66\times 10^6$ parameters for the encoder.

We trained the contrastive model for up to $450$ epochs with a batch size of $64$ and a learning rate of $1\times 10^{-3}$, and $2$ steps of gradient accumulation on one H100 GPU. We implement a specialized batching method to address issues with collating irregularly-sized meshes. We then trained the generative model for $1\times10^6$ epochs using a batch size of $2048$, a learning rate of $1\times 10^ {-5}$, and $2$ gradient accumulation steps on one H100 GPU. 

\begin{figure*}[ht!]
    \centering
    \includegraphics[width=0.95\linewidth]{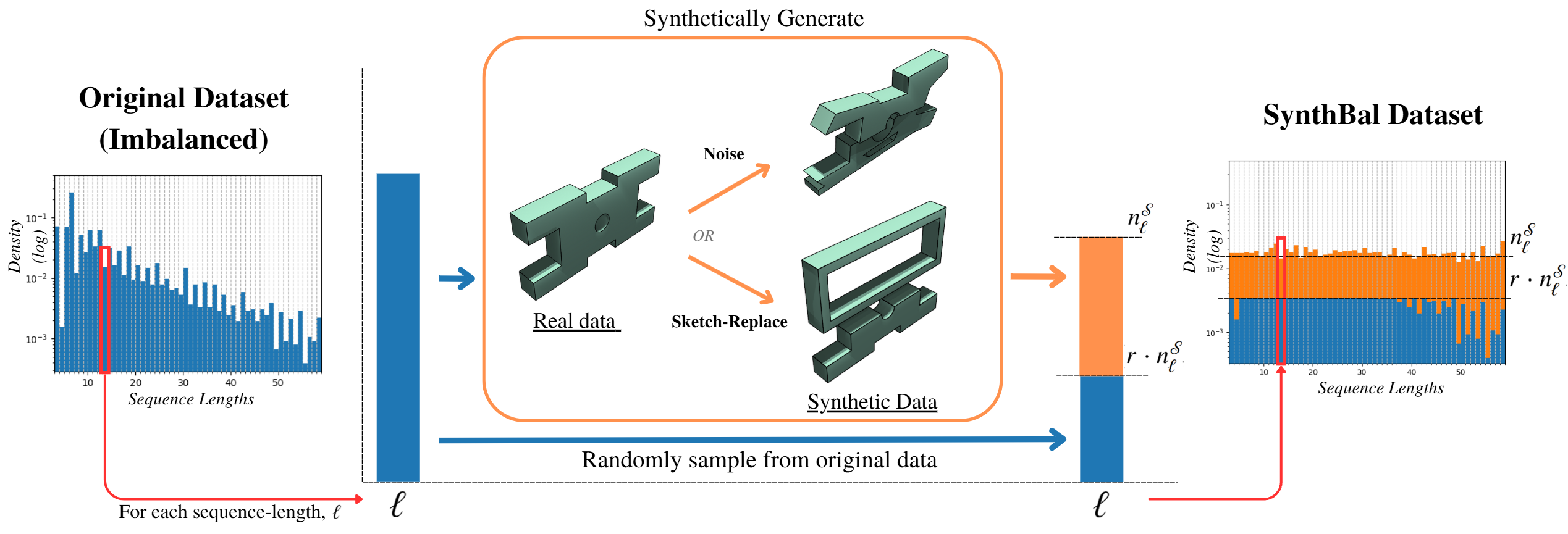}
    \caption{\textbf{SynthBal Generation Architecture.} We introduce a method to address sequence-length imbalance by generating synthetic CAD program data for under-represented sequence lengths. The sequence-length imbalance in the DeepCAD dataset is visualized (log-scale). }
    \label{fig:synthbal}
\end{figure*}

\section{Dataset} 
We extend the DeepCAD \cite{wu2021deepcad} dataset to 3D modalities for training and testing our models across all modalities. The DeepCAD dataset consists of $178,238$ CAD programs. Some of these programs do not compile into valid shapes (see Sec. \ref{sec:metrics}), and thus we remove them for training. This leaves us with $168,656$ CAD programs. These CAD programs range in complexity from sequence lengths of $3$ to $59$, the distribution of which is shown in Fig. \ref{fig:synthbal}.

Next, we use the CAD program dataset to create each of the modality specific datasets. To generate point clouds, we collect uniform samplings of 4096 points on mesh representations of the CAD programs. We generate the meshes through OpenCascade by first compiling the CAD program into a B-Rep and then converting it into a mesh. We used fine meshing options of $0.001$ linear deflection and $0.1$ angular deflection to maintain high resolution. We use the Trimesh Python package \cite{trimesh} to produce the uniform sampling of points. For each point, we also collect the surface normal from the mesh face that the point was sampled from. For the mesh dataset, we directly collected the vertices, vertex normals, and edges of the corresponding meshes. We calculate the vertex normals as the average of adjacent face normals, as implemented in Trimesh. 

We re-mesh the original meshes generated by OpenCascade to distribute the vertices uniformly across the mesh and to standardize the vertex count. This was done with the software MeshLab \cite{LocalChapterEvents:ItalChap:ItalianChapConf2008:129-136}. We chose a target vertex count of 2000 to 4000 vertices, similar in magnitude to the point cloud dataset. We first use three iterations of midpoint subdivision to increase sampling of the mesh surface, then we use five iterations of isotropic re-meshing \cite{10.1145/1057432.1057457} to improve uniform vertex distribution over the surface. The re-meshing process produced errors in some meshes, which we removed from the dataset. Our training dataset size is thus reduced from 152,530 to 136,384 meshes.

\section{Synthetic Balanced Dataset (SynthBal)}

The original DeepCAD dataset shows a significant data imbalance with regard to complexity, here defined as the sequence length of a CAD program. High-complexity CAD programs are underrepresented and low-complexity CAD programs are overrepresented: while the dataset contains programs of sequence length 3 to 59, sequence length of at most 6 constitute 42\% of the dataset, and sequence lengths of exactly 6 constitute 26\% (Fig. \ref{fig:synthbal}). The imbalance in the original dataset biases the performance towards low-complexity CAD programs (as shown in Fig. \ref{fig:reconstruction} and Tab.~\ref{tab:shape_encoding}). Prior works have also highlighted this challenge  \cite{jung2024contrastcad, wu2021deepcad}, and have experimented with dataset augmentation to address the scarcity of data at high complexity. However, they do not address the inherent imbalance in proportion, potentially leading to poorer performance for longer sequence lengths. We address this issue by adapting Ye-Bin et al \cite{yebin2024synaugexploitingsyntheticdata} into SynthBal, a new synthetic dataset augmentation scheme with an equal balance of CAD programs across sequence lengths. The scheme is visualized in Fig. \ref{fig:synthbal} and summarized in pseudocode in Alg. \ref{alg:synthbal}. 

We generate the synthetic dataset $\mathcal{S}$ of size $N^{\mathcal{S}}$ from a real dataset $\mathcal{R}$ as follows. Let $\mathcal{L}$ describe the set of existing sequence lengths in a dataset, and for any dataset $\mathcal{X}$, let $n^{\mathcal{X}}_{\ell}$ denote the number of datapoints that have sequence length $\ell$. Note that in an evenly balanced dataset, $n^{\mathcal{S}}_{\ell}=N^{\mathcal{S}} / |\mathcal{L}|$. For each sequence length, we first populate a ratio $r$ of the new dataset $\mathcal{S}$ using real data from $\mathcal{R}$, or as much as available. In other words, we move $\text{min}(r \cdot n^{\mathcal{S}}_{\ell}, n^{\mathcal{R}}_{\ell})$. We interpret $r$ as the limit of real data per sequence length. The rest of the data will be generated synthetically. We consider two augmentations as building blocks:
\vspace{0.5cm}
\begin{itemize}
    \item Noise: Add perturbations to ratio $p$ of sketch and extrude commands. We only perturb continuous values, ignoring discrete or “near-discrete” values (e.g., extrusion orientation is typically 0$^\circ$, +90$^\circ$, -90$^\circ$). Perturbations are uniformly sampled from the range $[- m \cdot 256, m \cdot 256]$ for a chosen magnitude $m$. We clip all perturbed values to the range $[1, 255]$.
    
    \item Replace-Sketch: Randomly select and replace some sketch operations in CAD program $\mathcal{C}_1$ with corresponding sketches of a randomly selected program $\mathcal{C}_2$. 

    We only pair data points from within the same training, validation, or test sets to prevent data leakage.
\end{itemize}

\begin{algorithm}[h]
\caption{SynthBal Dataset Generation}\label{alg:synthbal}
\textbf{Input:} Original Dataset $\mathcal{R}$

\textbf{Parameters:} Max ratio of real data $r$, Desired synthetic dataset size $N^\mathcal{S}$, Weighted choices of augmentations $\mathcal{A}$

\textbf{Output:} Balanced dataset $\mathcal{S}$

\For{Sequence length $\ell \in \mathcal{L}$}{
    $\mathcal{S}_\ell \gets \{\}$ \tcp*{Synthetic data of length $\ell$} 
    $n^{\mathcal{S}}_\ell \gets N^\mathcal{S}/|\mathcal{L}|$\;
    $n_r \gets min(r\cdot n_\ell^{\mathcal{S}}, n_\ell^{\mathcal{R}}$)\;
    $\mathcal{S}_\ell$ += \textit{sample_without_replace}($\mathcal{R_\ell}, n_r$) \;
    \While{$|\mathcal{S}_\ell| < n^{\mathcal{S}}_\ell$}{
        $\mathcal{C} \gets \textit{sample_from}(\mathcal{R_\ell})$\;
        $a \gets \textit{sample_from}(\mathcal{A})$\;
        $\mathcal{C} \gets a(\mathcal{C})$\;
        \eIf{$\textit{is_valid}(\mathcal{C})$}{
            $\mathcal{S}_\ell += \mathcal{C}$\;
        }{
            skip
        }
    }
    $\mathcal{S} += \mathcal{S}_\ell$
  }
\end{algorithm}

We generate augmentations in the following way. For a given CAD model, randomly select between the following options: with $40\%$ probability, apply a large noise ($m=0.07$, $p=0.6$) or with 60\% chance, apply small noise ($m=0.02$, $p=0.8$) then a Replace-Sketch. We then check if the augmentation produced a valid CAD object. We perform two checks: first, if it compiles into a CAD object, and second, if it is free of self-intersections. As opposed to DeepCAD and ContrastCAD, which perform augmentations during training and do not validate them, it is computationally feasible for us to validate our augmentations because they are generated outside of training.

Another advantage of our approach is that we can create balanced datasets of significantly larger sizes, addressing the relative lack of datapoints in the DeepCAD dataset. We produce the \textbf{SynthBal dataset}  ($N^{\mathcal{S}}=170,000$ and $r=0.2$), which is equivalent in size to the original DeepCAD dataset. 16.7\% of this dataset is copied from the original, and 83.3\% is synthetic. We also produce a larger \textbf{SynthBal-1M dataset} ($N^{\mathcal{S}}=1,000,000$ and $r=0.8$), for which 13.9\% is copied from the original dataset and 86.1\% is synthetic.
CAD programs associated with the original train, test, validation sets are split accordingly for the SynthBal sets. This prevents data leakage when evaluating on the original dataset. While we use the SynthBal dataset during training, we perform all evaluations on the original dataset for consistent comparison to existing methods. 

We performed training following the method introduced by SYNAuG \cite{yebin2024synaugexploitingsyntheticdata}, to address image classifier training on long-tailed class-imbalanced datasets. It describes first training on a synthetically balanced dataset, then fine-tuning the last layer on a "reduction-balanced" dataset of purely real data. This reduction-balanced dataset addresses the imbalance in the original dataset by randomly removing data from overrepresented classes as opposed to adding synthetic data to underrepresented classes. We adapt this approach to training our CAD program autoencoder. We first train on the SynthBal dataset and then fine-tune on a reduction-balanced DeepCAD dataset with $4,503$ CAD models. We allow all autoencoder parameters to be trained during fine-tuning.

For the synthetic training phase, we use the SynthBal dataset and trained for $1000$ epochs with a learning rate $1\times10^{-3}$, warmup of $2000$ steps, and a batch size of $512$. During fine-tuning, we use the reduction-balanced dataset and train for $100$ epochs using a learning rate of $1\times10^{-5}$, $200$ warmup steps, and a batch size of $512$. The autoencoder architecture is described in Section~\ref{sec:method_1}.

We call the final fine-tuned autoencoder the \textbf{SynthBalFT autoencoder}. Without fine-tuning, we call it the \textbf{SynthBal autoencoder}. We refer to all autoencoders trained with synthetic data \textbf{Synthetic autoencoders}, which includes the SynthBalFT and SynthBal-1MFT models. When trained on the original dataset, we call it the \textbf{Original autoencoder}.


\section{Metrics} \label{sec:metrics}
All metrics reported in this paper are evaluated on the original DeepCAD dataset for consistency of comparison.

\textbf{Sequence-Length Normalization:}
Most existing studies report performance metrics averaged across the entire dataset. When the dataset is imbalanced, the resulting metric primarily reflects the performance of the overrepresented classes and hides the performance of the underrepresented classes. 

We introduce a new normalized metric to address this phenomenon. For each metric $m$ applied to some dataset $\mathcal{X}$, we report two values: the metric directly applied to the set of all CAD programs $m(\mathcal{X})$, and the metric normalized across sequence lengths 
\begin{equation}
    m^{\text{SL}}(\mathcal{X}) := \frac{1}{|\mathcal{L}|} \sum_{\ell \in \mathcal{L}} m(\mathcal{X}_{\ell}). 
\end{equation}

Here, $\mathcal{X}_\ell$ refers to the datapoints of sequence length $\ell$ in $\mathcal{X}$. In simpler words, we average across the metric computed for each sequence length. We show that these values emphasize the performance of the models at high sequence lengths, which are underrepresented in the dataset.

\textbf{Reconstruction Accuracy:} We use reconstruction accuracy in two contexts: first to assess the ability of our autoencoders to reconstruct an input CAD program, and second to assess the ability of our conditional generative models to reconstruct a CAD program that matches an input geometry. Since our generative models are stochastic, we evaluate the accuracy based on a single generated sample. We use the same metrics proposed by DeepCAD \cite{wu2021deepcad} and introduce a new metric, intersect over union:

\begin{itemize}
    \item Command Accuracy $(\mu_{\text{cmd}})$: The ratio of correctly predicted commands. 
    \begin{align*}
    \mu_{\text{cmd}} = \frac{1}{N} \sum^N_{k=1} \mathbb{I}[t_k = \hat{t}_k]
    \end{align*}
        
    \item Parameter Accuracy ($\mu_{\text{param}}$): The ratio of parameters correctly predicted within tolerance. 
    \begin{align*}
    \mu_{\text{param}} = \frac{1}{T} \sum^N_{k=1} \sum^{|\hat{p}_k|}_{l=1} \mathbb{I}[|p_{k,l}-\hat{p}_{k,l}| < \eta] \mathbb{I}[t_k = \hat{t}_k]
    \end{align*}
    
    where $T=\sum^N_{k=1} \mathbb{I}[t_k=\hat{t}_k] |p_k|$ is the number of correctly predicted parameters, $|p_k|$ is the number of parameter types, and $\eta$ is the tolerance for accuracy.
    
    \item Invalid Ratio (IR): Not all CAD programs can be compiled into a valid geometry. Consistent with DeepCAD, we classify a CAD program as invalid if OpenCascade fails to compile it into a B-Rep or if that B-Rep fails to be converted into a pointcloud.
    
    \item Chamfer Distance ($M_{CD}$): The similarity between the surfaces of validly \textit{compiled} CAD geometries. For two compiled geometries, we calculate the chamfer distance between 2000 points uniformly sampled from each geometry. We report the median value across all chamfer distances in the test set.
    
    \item Intersect over Union (IoU): The similarity between the volumes of validly \textit{compiled} CAD geometries. For two geometries $G_1, G_2$, this value is $\frac{G_1 \cap G_2}{G_1 \cup G_2}$. To accurately calculate this value, we first optimally align normalized versions of these geometries using the method introduced in \cite{dorisIou}.
\end{itemize}

\textbf{Retrieval Top-N Accuracy}: Top-N accuracy evaluates the alignment quality between modalities by measuring the proximity of corresponding CAD and geometry embeddings. During retrieval of a given geometry, we return the closest $N$ CAD programs from a library of $R_B$ CAD programs, with proximity defined by cosine similarity of embeddings $p(e_1, e_2) = e_1 \cdot e_2 / ||e_1|| ||e_2||$. We evaluate the Top-1 retrieval accuracy over library batch sizes of $10$, $128$, $1024$, and $2048$ bootstrapped from the full test set of $7,616$ CAD programs. For each batch size, we average across 100 evaluations to reduce statistical deviation.
    
\textbf{Generation Quality}: 
We assess whether the generative model can produce both "diverse" and "realistic" CAD programs by evaluating the statistical similarity of a batch of generated CAD B-Reps with respect to a batch of real CAD B-Reps. We report the metrics described by Wu et al \cite{wu2021deepcad} as standard metrics of comparison to related works. Specifically, we report the Jensen-Shannon divergence (JSD), which assesses generation "quality" as the statistical distance between the real and generated sets of CAD geometries; coverage (COV), which assesses the "diversity" of generation as the percentage of the real dataset that matches objects in a generated set; and the minimum matching distance (MMD), which assesses generation "fidelity" as the similarity of corresponding objects in the real and generated sets.

\textbf{Comparing Accuracy}: We compare accuracy metrics by relative improvements in error: $p_1$ improves upon $p_2$ by $\frac{p_1 - p_2}{1-p_2}$. For example, an accuracy of $0.995$ improves upon an accuracy of $0.990$ by 50\%, as does $0.95$ upon $0.90$. We do so to better reflect improvements by order of magnitude.


\section{Results}
\subsection{Synthetic Balancing Improves Autoencoder Reconstruction}

\newcolumntype{x}[1]{%
>{\centering\hspace{0pt}}p{#1}}%

\begin{table*}
    \centering
    \caption{\textbf{Sequence Length Normalization of Metrics Reduces Bias Towards Overrepresented Sequence Lengths.} Demonstrated with the DeepCAD autoencoder \cite{wu2021deepcad}. Performance metrics vary heavily across sequence lengths, but the unnormalized metrics primarily reflect performance around sequence lengths 5-15 (we underline values closest to the unnormalized metrics). }

    \begin{NiceTabular}{x{2.5cm} p{1.5cm} | p{1.3cm} p{1.3cm} p{1.3cm} p{1.3cm} c}
    
    \toprule
        Sequence Length    & Ratio (\%)& $\mu_{\text{cmd}} (\uparrow)$ & $\mu_{\text{param}} (\uparrow)$ & $M_{\text{CD}} (\downarrow)$ &  $IoU (\uparrow)$ & $\text{IR} (\downarrow)$ \\
    \midrule
        5  & 6.99 & $100.0$ & $99.52$ & $\underline{0.754}$ & 99.8& $0.907$ \\ 
        15  & 1.64 & $\underline{99.81}$ & $\underline{97.58}$ & $0.857$ &\underline{92.1}& $\underline{2.02}$ \\ 
        25  & 0.773 & $98.78$ & $93.62$ & $2.12$ & 79.4& $6.38$ \\ 
        35  & 0.328 & $96.20$ & $88.81$ & $4.54$ & 55.8& $15.85$ \\ 
        45  & 0.197 & $93.64$ & $84.59$ & $11.3$ & 48.9& $20.31$ \\ 
        55  & 0.0393 & $91.13$ & $77.28$ & $21.4$ & 48.7 &$25.56$ \\ 
    \toprule
        \multicolumn{2}{c}{Metric}\\
    \midrule 
        \multicolumn{2}{c}{No Normalization ($m(\mathcal{X})$)} & $ 99.36 $ & 97.59 & 0.783 & 93.7& 3.44 \\    
        \multicolumn{2}{c}{Normalized, Ours ($m^{\text{SL}}(\mathcal{X})$)}  & $96.39$ & $ 90.12$ & $7.82$ &72.6& $ 12.22$ \\  

   \bottomrule 
    \end{NiceTabular}
    \label{tab:imbalance}
\end{table*}

\begin{table*}
    \centering
    \caption{\textbf{Synthetically Balanced Data Improves Autoencoder Reconstruction Performance}. Arrows indicate whether higher or lower is better. Bold numbers represent the overall best performance. Pretrained models from literature were downloaded then re-evaluated on the test set. * indicates our GenCAD-3D models.}

    \begin{NiceTabular}{l| c@{\hskip 5pt}c@{\hskip 5pt}c@{\hskip 5pt}c@{\hskip 5pt}c@{\hskip 5pt} |[tikz=dashed] c@{\hskip 5pt}c@{\hskip 5pt}c@{\hskip 5pt}c@{\hskip 5pt}c@{\hskip 5pt}}
    \toprule
        Model    & $\mu_{\text{cmd}} (\uparrow)$ & $\mu_{\text{param}} (\uparrow)$ & $M_{\text{CD}} (\downarrow)$ &  $IoU (\uparrow)$ & $\text{IR} (\downarrow)$ & $\mu_{\text{cmd}}^{\text{SL}} (\uparrow)$ & $\mu_{\text{param}}^{\text{SL}} (\uparrow)$ & $M_{\text{CD}}^{\text{SL}} (\downarrow)$ & $IoU^{\text{SL}} (\uparrow)$ & $IR^{\text{SL}} (\downarrow)$\\
    \midrule
        *SynthBal-1M FT & $99.66$ & $\textbf{98.61}$ & $0.734$ & \textbf{96.5}&$\textbf{0.807}$ & $98.52$ &$\boldsymbol{94.72}$  & $ \textbf{2.00}$ & \textbf{83.0}& $3.57 $\\  
        *SynthBal-1M  & $99.65$ & $98.59$ & $0.736$ & 96.1& $0.969$ & $98.47$ &$94.59$  & $ 2.54$ & 79.8& $4.66 $\\  
        *SynthBal FT  & $99.70$ & $98.34$ & $0.757$ & 95.7& $0.845$ & $\boldsymbol{98.59}$ & $93.85$  & $ 2.87$ & 79.2&$3.56 $\\    
        *SynthBal & $99.68$ & $98.22$ & $0.767$ & 95.1& $0.845$ & $98.57$ & $93.56$  & $2.67$ & 76.5&$ \textbf{3.36}$\\ 

        \hdashline

        *Original & $99.51$ & $97.78$ & $0.772$ &94.2&  $3.32$ & $97.00$ & $91.11$ & $5.84$& 69.1& $ 11.96$\\  

        ContrastCAD+RRE \cite{jung2024contrastcad} & $\boldsymbol{99.76}$ & $98.58$ & $\boldsymbol{0.731}$ & 95.8& $2.53$ & $98.24$ & $93.02$ & $4.69$& 73.6&$ 8.49$\\   

        DeepCAD \cite{wu2021deepcad} & $ 99.36 $ & 97.59 & 0.783 & 93.7& 3.44 & $96.39$ & $ 90.12$ & $7.82$ & 72.6&$ 12.22$\\ 

   \bottomrule 
    \end{NiceTabular}
    \label{tab:shape_encoding}
\end{table*}

\begin{figure}[!h]
    \centering
    \includegraphics[width=0.95\linewidth]{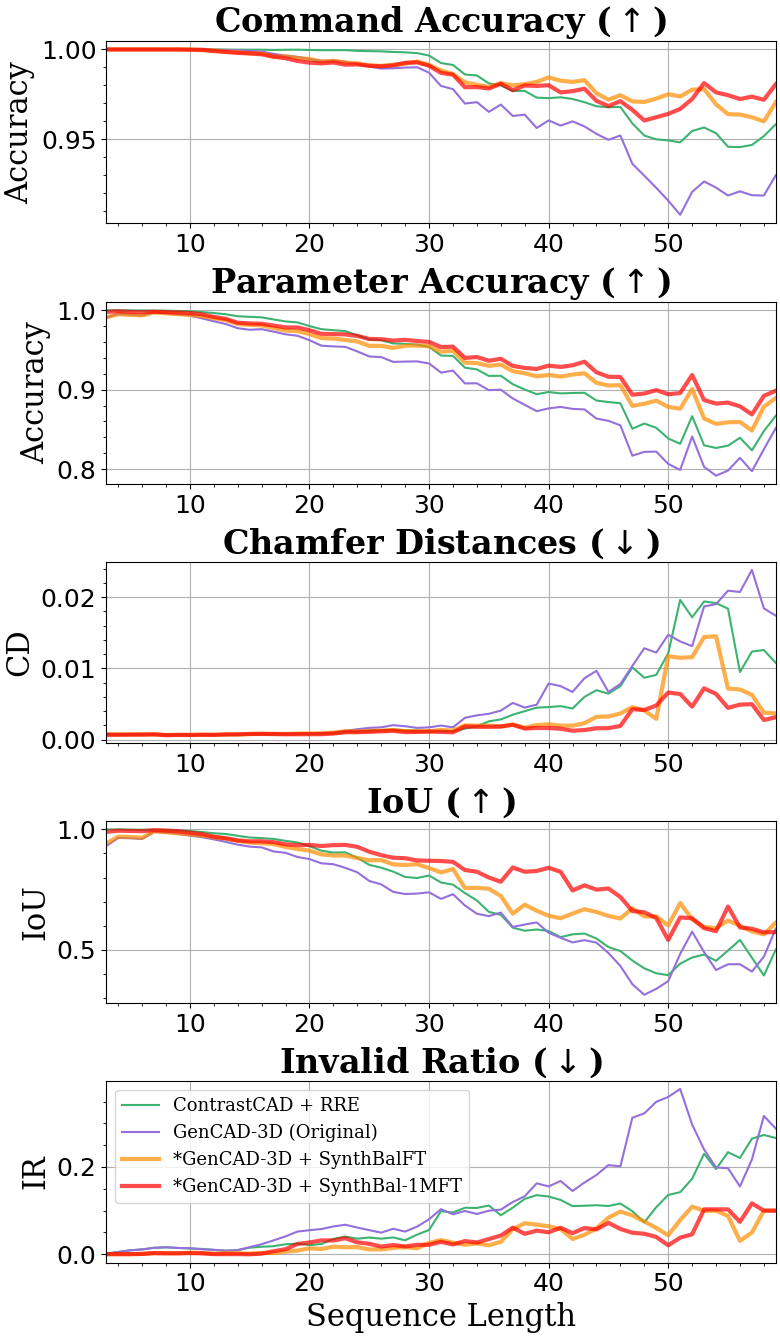}
        \caption{\textbf{Reconstruction Metrics Across Sequence Lengths.} Our SynthBalFT autoencoders (*) improve reconstruction of high sequence length CAD programs. (Smoothed with uniform filter, size=5)}
    \label{fig:reconstruction}
\end{figure}

Our new sequence length normalization metric is better at reflecting performance across sequence lengths. As seen in Fig. \ref{fig:reconstruction} and Tab. \ref{tab:imbalance}, there are large variations in performance across sequence lengths and a noticeable reduction in performance at longer, underrepresented sequence lengths. For example between the sequence length of 55 and 5 with 0.0393\% and 6.99\% representations respectively, the DeepCAD autoencoder shows a 9.8\% absolute difference in command accuracy and a 24.6\% absolute difference in invalid ratio. We also see that the original, unnormalized metric primarily reflects performance at short, overrepresented sequence lengths around 5 to 15. Our normalization metric is thus better at capturing overall performance.

Our synthetic autoencoders\footnote{We use "synthetic model" to refer broadly to any model trained with synthetic data, including the SynthBal and SynthBal-1M autoencoders.} significantly improved performance, especially at high sequence lengths (Tab. \ref{tab:shape_encoding}). 
All of our synthetic autoencoders outperformed the baselines in all sequence-length normalized metrics: our SynthBal-1MFT model significantly reduced error in command accuracy (16\%), parameter accuracy (24\%), chamfer distance (57\%), IoU (36\%), and invalid-ratio (58\%) compared to the next-best ContrastCAD+RRE model (RRE refers to their Random Replace and Extrude metric, see Sec. \ref{sec:synthetic-aug}). 
Across unnormalized metrics, our SynthBal-1MFT model also showed improved performance compared to ContrastCAD in parameter accuracy, IoU, and invalid ratio. We also see that fine-tuning consistently improved our models' performances in command accuracy, parameter accuracy, and IoU for both the SynthBal and SynthBal-1M autoencoders.

We see that artificially increasing the size of our dataset with synthetic data improves performance: our SynthBal-1MFT autoencoder trained on 1 million datapoints outperforms our SynthBalFT autoencoder trained on 170 thousand datapoints in parameter accuracy, chamfer distance, and IoU, while performing similarly in invalid ratio (within 0.3\%).

Nevertheless, we also see that dataset balancing was essential to improving performance: our SynthBalFT model outperformed both ContrastCAD and our original autoencoder in all sequence-length normalized metrics despite being trained on the same dataset size.

Our synthetic autoencoders significantly improved the invalid ratio even in the unnormalized metrics. Compared to our original GenCAD-3D model, our SynthBal-1MFT model showed a 76\% relative improvement in unnormalized invalid ratio and a 70\% relative improvement in normalized invalid ratio. Compared to the next-best ContrastCAD+RRE model, our SynthBal-1MFT model showed a 68\% relative improvement in unnormalized invalid ratio and a 58\% relative improvement in normalized invalid ratio. We hypothesize that by enforcing valid CAD program augmentations, particularly at high sequence lengths, we allowed the optimizer to better discover the manifold of valid CAD sequences. 

When looking at individual sequence lengths (Fig. \ref{fig:reconstruction}), we more clearly see how our synthetic models improve performance at high sequence lengths. Compared to our original GenCAD-3D model, our synthetic models consistently improved performance across all sequence lengths. Compared to the ContrastCAD+RRE model, we see that our SynthBal models significantly improved performance at high sequence lengths, starting around sequence length 33. However, they did not perform as well in command and parameter accuracy at low sequence lengths. We believe this is caused by the ContrastCAD+RRE model overfitting to low sequence lengths. We also see that our SynthBal-1MFT autoencoder performed better than the SynthBalFT autoencoder at high sequence lengths, particularly for parameter accuracy and chamfer distance. 

Overall, training the autoencoder on the SynthBal dataset showed considerable improvements in performance as a result of balancing, increased dataset size, and validity checking. It is noteworthy that performance was still strong even though the SynthBal dataset consisted mostly of synthetic data, whereas the test set consisted entirely of real data. This suggests that synthetic data can be feasibly used as data augmentation to transfer to real use cases. We highlight that while SynthBal was tailored to address imbalance in CAD datasets, the overall framework may be applicable to any engineering dataset facing issues with balancing and small size.

\begin{figure}
    \centering
    \includegraphics[width=1\linewidth]{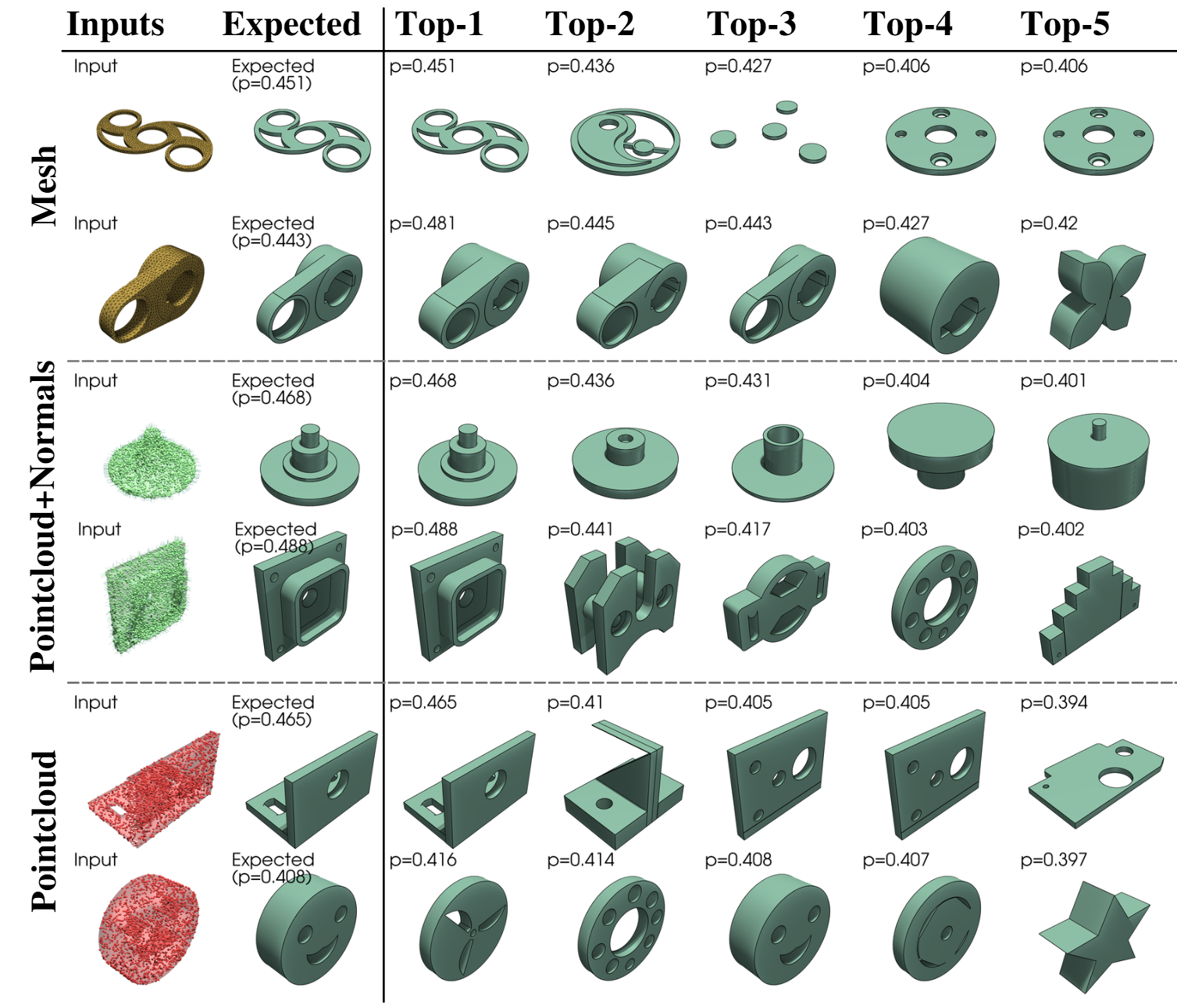}
    \caption{\textbf{Selected Top-5 Retrieval Results from Multimodal Inputs.} Not only is the latent space capable of accurately identifying the closest CAD object, but it is also able to find similar CAD objects. We use cosine similarity $p$ between CAD and geometry embeddings to determine ranking.}
    \label{fig:pcn-retrieval}
\end{figure}

\subsection{Aligned Latent Space Enables Multimodal Retrieval from a CAD Library}

We evaluate the performance of multimodal latent embeddings for retrieving CAD programs from a library, as shown in Table \ref{tab-top1-geom_synth}. We trained the contrastive step of these models using the synthetic autoencoders on the Original training dataset (we justify these choices through ablation studies in section \ref{sec:ablation}). 

We show that our contrastive models show strong performance in retrieval: for moderate library batch sizes (128), our contrastive models trained with SynthBal-1MFT were able to retrieve the exact matching CAD program at least 96\% of the time across modalities, whereas for large library sizes (2048), we can retrieve an exact match at least 81\% of the time. We also visualize the top-5 retrievals in Fig. \ref{fig:pcn-retrieval} for selected examples. Even when retrieval does not obtain an exact top-1 match, the correct CAD usually appears within the top-5 matches, making it easy for a user to review. We also see that retrieval returns not only the exact corresponding CAD program but a selection of geometrically similar models, providing a user with a diverse set of options to select from. For more examples of retrieval, see our project site \url{gencad3d.github.io}.

Between modalities, we see that the models that incorporate mesh information had the best performance, with the point cloud+normal model consistently showing at least a 7\% improvement in retrieval compared to the pure-point cloud model from a library of 2048. The mesh model consistently shows the best performance across all retrieval library sizes and across all autoencoder metrics: with the original autoencoder, it showed a 26\% and 16\% improvement in retrieval compared to the pure-point cloud and the point cloud + normal models, respectively, in retrieval from 2048. When trained with the SynthBal-1MFT autoencoder, the mesh model showed a 11\% and 4\% improvement in retrieval compared to the pure-point cloud and point cloud + normal models, respectively. 

Between autoencoders, we see that all of our synthetic models outperformed our baseline model trained with the original autoencoder in retrieval, confirming that our SynthBal strategy improves performance in downstream tasks. However, the amount of improvement gained by switching to our synthetic models diminished for the mesh modality, with our SynthBal-1MFT model providing 25\%, 22\%, and 7\% improvements in retrieval from 2048 for the pointcloud, pointcloud + normal, and mesh models, respectively. We also note that only in the mesh modality, the SynthBal-1MFT did not significantly improve retrieval compared to SynthBalFT, with a 0.54\% and 1.6\% improvement in retrieval from 2048 and 1024 and worsened retrieval from 128. 

Nevertheless, we see that artificially increasing dataset size using SynthBal-1M during autoencoder training consistently improves downstream retrieval performance for the point cloud and point cloud + normals modalities.

\begin{table*}[h]
    \centering
        \caption{\textbf{Top-1 Multimodal Retrieval Scores.} GenCAD-3D shows accurate retrieval even through large libraries. Training with the SynthBal autoencoder consistently improves performance. Bold indicates the best performance within each modality. The contrastive steps were all trained with the original dataset.}
    \begin{tabular}{cc|rrrr}
    \toprule
    Modality & Autoencoder & $R_B=10$ & $R_B=128$ & $R_B=1024$ & $R_b=2048$ \\
    \midrule
    
    Point Cloud & Original &  $99.4 $ & $95.6 $    & $81.1 $  & $72.3 $ \\
    Point Cloud & SynthBalFT &  $99.4$ & $96.2$    & $83.6$  & $75.6$ \\
    Point Cloud & SynthBal-1MFT &  $99.8$ & $\textbf{97.1}$    & $\textbf{86.1} $  & $\textbf{79.4}$ \\
    \hline

    PC+N & Original  &  $100.0 $   & $96.8 $    & $83.6 $     & $75.5 $ \\
    PC+N & SynthBalFT  &  $99.8$   & $97.1$    & $85.1$     & $77.5$ \\
    PC+N & SynthBal-1MFT  &  $100.0 $   & $\textbf{97.2}$    & $\textbf{87.4} $     & $\textbf{80.6} $ \\
    \hline

    Mesh   & Original & 99.5 & 97.4  & 87.0   & 80.2\\
    Mesh    & SynthBalFT &  99.7 & \textbf{97.8}  & 87.7 & 81.5\\
    Mesh    & SynthBal-1MFT & 99.7& 97.2  & \textbf{87.9} & $\textbf{81.6}$\\
    \hline

    Random  & - &$10.2 $ &$ 0.80 $ & $ 0.082 $ & $0.048 $\\
    \bottomrule \\
    
    \end{tabular}
    \label{tab-top1-geom_synth}
\end{table*}


\begin{table*}[h]
        \centering
        \caption{\textbf{Quality and Accuracy of Multimodal Generative Models.} Training with the synthetically balanced autoencoders and modalities with mesh information improves conditional reconstruction accuracy. The generative steps of all models were trained with the original dataset. * indicates our models.  "-" indicates that the metric is not relevant. Values from models from literature were re-evaluated when pretrained models were available and copied from associated papers when not. }
    \begin{NiceTabular}{c@{\hskip 0pt}c@{\hskip 3pt}c@{\hskip 2pt}c@{\hskip 1pt}|c@{\hskip 1pt}c@{\hskip 2pt}c@{\hskip 2pt}c@{\hskip 2pt}c@{\hskip 1pt}|[tikz=dashed] c@{\hskip 1pt}c@{\hskip 2pt}c@{\hskip 2pt}}
    \toprule
    Model & Modality & Autoencoder  & \makecell{Contrastive \\Dataset} & $\mu_{\text{cmd}}^{\text{SL}} {\scriptstyle (\uparrow)}$ & $\mu_{\text{param}}^{\text{SL}} {\scriptstyle (\uparrow)}$ & $M_{\text{CD}}^{\text{SL}} {\scriptstyle (\downarrow)}$ &$ IoU^{\text{SL}} {\scriptstyle (\uparrow)}$& $IR^{\text{SL}} {\scriptstyle (\downarrow)}$ & COV${\scriptstyle (\uparrow)}$  & MMD${\scriptstyle (\downarrow)}$  & JSD${\scriptstyle (\downarrow)}$      \\ 
    
    \midrule
    
    *GenCAD-3D  & Point Cloud & Original  & Original & 42.45  &  42.02    &  51.16 &  37.6 & 17.66 & $\textbf{83.6}$ & $1.29 $ & 3.38 \\ 
    *GenCAD-3D   &   Point Cloud & SynthBalFT  &  Original & 45.33 & 45.94 & 62.55 & 37.2 & \textbf{15.10}  & $82.1 $ & 1.28 & $ 3.61$ \\ 
    *GenCAD-3D   &   Point Cloud & SynthBal-1MFT  &  Original &   52.83 & 50.72 & 52.10  & 45.8 & 21.03  & 83.1 & \textbf{1.20} & \textbf{3.34} \\ 
    *GenCAD-3D   &   Point Cloud & SynthBal-1MFT  &  SynthBal &  \textbf{53.57}  & \textbf{51.42} & \textbf{48.99} & \textbf{46.8} &  21.81  & 83.4 & 1.23 & 3.44 \\
         
    \hdashline

    *GenCAD-3D  & PC+N & Original   &  Original &    48.35  &   43.33  &     59.56    &  38.8 & 22.56 & 82.9 & $ 1.26$ & $ \textbf{3.34}$ \\ 
    *GenCAD-3D  & PC+N & SynthBalFT   &  Original &  50.99  & 47.31   & 56.65    &  41.7 & \textbf{18.34} &  $82.4$ & $1.27$ & $ 3.50$ \\ 
    *GenCAD-3D  & PC+N & SynthBal-1MFT   & Original & 55.73   & 50.62   & 46.03     & \textbf{44.8}&  22.09 & \textbf{83.3 }& $\textbf{1.21}$ & $ 3.49$ \\ 
    *GenCAD-3D  & PC+N & SynthBal-1MFT   &  SynthBal &\textbf{56.42}  &  \textbf{51.15} &   \textbf{45.80 }&  44.6 & 25.50 & 82.6 & \textbf{1.21} & 3.55 \\ 

    \hdashline
 
    *GenCAD-3D  & Mesh    & Original  & Original & 52.23  &  44.64   &   50.47  & 39.6 & 26.05 & 83.6 & 1.25 & \textbf{3.40} \\ 
    *GenCAD-3D   & Mesh  & SynthBalFT  & Original &  55.98    & 48.63  & 66.27    & 41.4 &  $\textbf{20.31}$ & 83.1 & 1.28 & 3.65 \\ 
    *GenCAD-3D   & Mesh  & SynthBal-1MFT  & Original &  60.10     & 52.18    & \textbf{49.68 }  & \textbf{43.8} & 25.26 & 83.7 & \textbf{1.18} &  3.49 \\ 
    *GenCAD-3D   & Mesh & SynthBal-1MFT  &  SynthBal & \textbf{60.48} & \textbf{52.64 }&  55.82  & 43.7 & 23.63 & \textbf{84.0} & \textbf{1.18} & 3.47 \\ 
         
    \hline

    DeepCAD \cite{wu2021deepcad} & Uncond. &-& -&-&- & - & -& -& 78.1& 1.45 & 3.76 \\
    SkexGen \cite{xu2022skexgen}& Uncond.   &-&- &- & -& -  & - &-& 83.6 & 1.48 & 0.81 \\
    ContrastCAD \cite{jung2024contrastcad} & Uncond. &-&- &- &-&- & - & -& 78.9 & 1.44 & 3.67 \\
    BrepGen \cite{xu2024brepgen} & Uncond.  &-&- &-&- &- & -  &-& 78.2 & 1.02 & 0.90 \\ 
    \bottomrule 
    \end{NiceTabular}

      \label{tab:generation_synth}
\end{table*}

\subsection{Generative Architecture Enables Multimodal Reconstruction and Generation}

\begin{figure}[!h]
    \centering
    \includegraphics[width=0.98\linewidth]{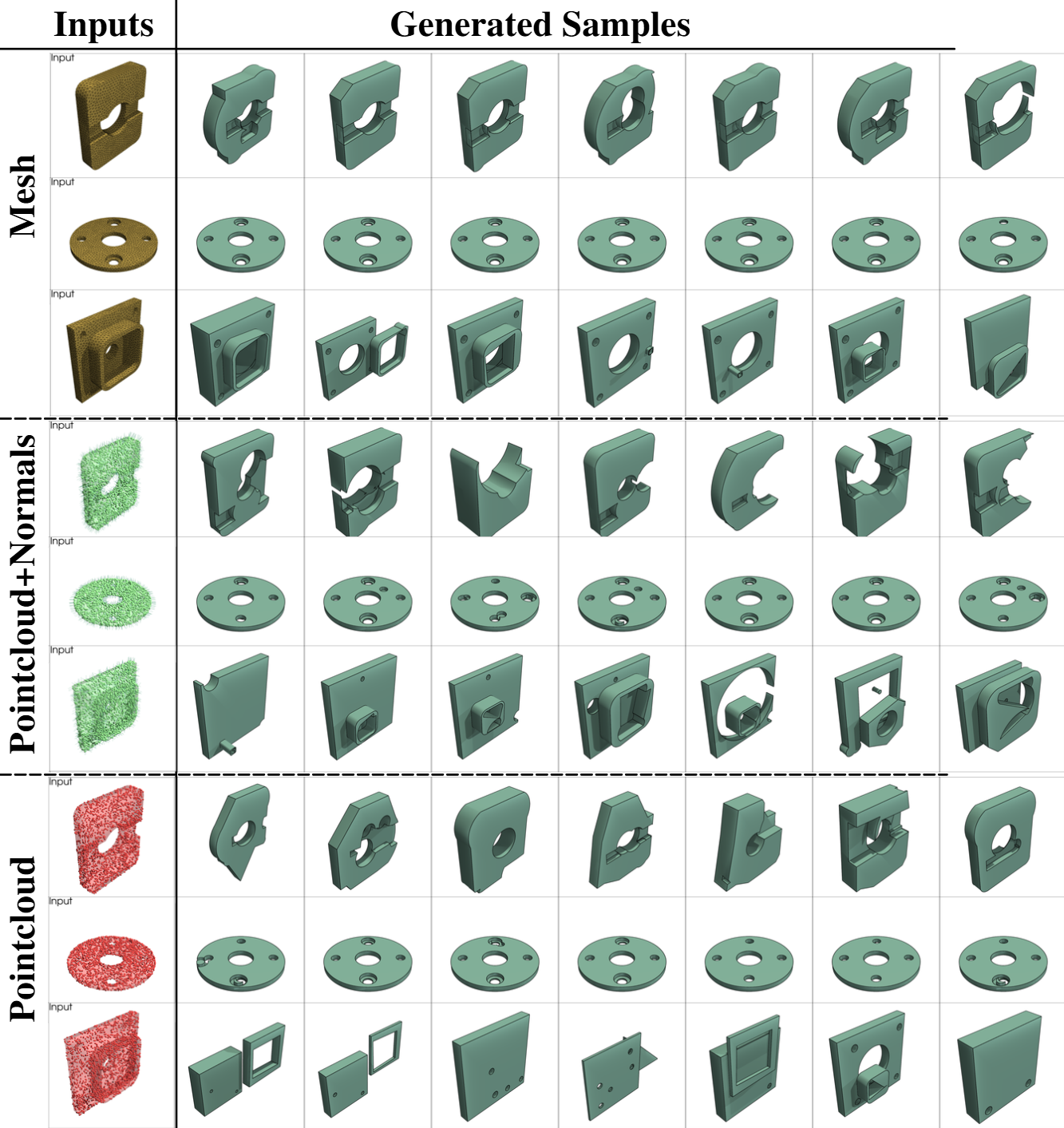}
    \caption{\textbf{Selected Generation Results from Multimodal Inputs.} GenCAD-3D is able to create CAD programs similar to the input while also producing variation.}
    \label{fig:generation}
\end{figure}

We look at the conditional reconstruction and generative capabilities of our generative model, reported in Table \ref{tab:generation_synth}. The generative training step of all models used the Original dataset (ablation studies justifying these choices are in Sec. \ref{sec:ablation}). While a variety of conditional reverse-engineering or generative models exist in literature \cite{khan2024cad, xu2024cad, dupont2025transcad, uy2022point2cyl}, they either do not release their code, report different metrics, or use inherently different representations of CAD programs, preventing us from quantitatively comparing our results. We thus only report unconditional generative models for quantitative comparison.

For conditional reconstruction, we assess how accurately the generative model can reproduce CAD programs that match a geometric input. We find that the reconstruction accuracy is relatively low compared to that from the autoencoders, with a maximum command accuracy of 60.48\% compared to the baseline autoencoder reconstruction accuracy of 99.36\%. We expect accuracy to be lower because the generative diffusion model inherently incorporates stochasticity to encourage variation in the outputs. In other words, these reconstruction metrics assessing similarity (command accuracy, parameter accuracy, chamfer distance, and IoU) should not be taken as absolute reflections of quality. Nevertheless, we can use them to show relative trends in performance. Overall, these results suggest that our model tends to produce CAD programs that are dissimilar from the original geometry. 

We also observe that the invalid ratio is relatively large in our generative models, with an invalid ratio of at least 15\%, especially compared to the low invalid ratio of our Synthetic autoencoders (maximum 4.66\%, Tab. \ref{tab:shape_encoding}). We suggest the following explanation: denoting the set of embeddings associated with some dataset $\mathcal{X}$ as $Z_{\mathcal{X}}$, this may suggest that a) our CAD program autoencoder is finding a latent manifold that captures valid CAD programs when nearby $Z_{\mathcal{X}}$, but it captures invalid CAD programs when farther from $Z_{\mathcal{X}}$; and b) our diffusion model is generating embeddings that are farther from $Z_{\mathcal{X}}$. There is no noticeable trend across the choice of autoencoder, suggesting that the issue does not lie in dataset choice. We believe that constraining the latent CAD program manifold to valid programs, such as through negative data \cite{regenwetter2024constraininggenerativemodelsengineering}, during autoencoder training may improve the invalid ratio.

Between the Synthetic and Original autoencoders, we see that training the generation model with Synthetic autoencoders significantly improved performance, with minimum improvements of 14\% in command accuracy, 13\% in parameter accuracy, and 7\% in IoU across all modalities. We did not notice trends in chamfer distance and invalid ratio. 

We also find that we achieve slightly better reconstructions results when training with the SynthBal dataset during the contrastive step (full ablations are in Sec. \ref{sec:ablation}), providing a minimum 1\% improvement to command accuracy and 1\% improvement to parameter accuracy across all modalities. There was no consistent trend among chamfer distance, IoU, or invalid ratio.

Between the SynthBal-1MFT and SynthBalFT autoencoders, we see that the SynthBal-1MFT models produced better reconstruction results, with a minimum improvement of 9\% in command accuracy, 6\% in parameter accuracy, 19\% in chamfer distance, and 4\% in IoU across all modalities. However, we also see that the SynthBal-1MFT models consistently had higher invalid ratios. Nevertheless, generating new CAD programs is a computationally cheap task, and so the heightened invalid ratio is not of strong concern. These results suggest that increasing dataset size through SynthBal is beneficial to conditional reconstruction.

Overall, these results confirm the effectiveness of our SynthBal strategy for balancing and enlarging datasets in improving conditional reconstruction performance.

Between modalities, we see that our mesh model produced better conditional reconstructions. Our SynthBal-1MFT mesh models produced the highest command and parameter accuracies (+9\% and +3\% respectively from point cloud + normals, and +15\% and +3\% respectively from point cloud). However, the mesh model also tended to produce higher invalid ratios, with the worst invalid ratio of 26\%. We found no consistent trends in chamfer distance nor IoU. Overall, these results suggest that the mesh models produce more faithful reconstructions.

For generation, we assess how well our model can produce a "diverse" set of "realistic" CAD programs and compare them to models from literature. Our models performed better in all metrics as compared to unconditional generation from DeepCAD and ContrastCAD, suggesting that it produce CAD programs with the same level of diversity and realism as those models. Between autoencoder models, we observe that training our generation model with the Synthetic autoencoders slightly worsens performance in coverage (maximum 1.5\% decrease) and JSD (maximum 0.3\% increase), suggesting a reduction in diversity and quality. We believe that this reduction in performance occurs in part because these metrics are evaluated with respect to batches of CAD programs that are randomly sampled from the original dataset and thus share the same imbalances in complex CAD program representation. We believe that finding a method to generate balanced batches of real CAD programs is essential to creating an unbiased metric.

We also show a selection of generated samples in Fig. \ref{fig:generation}. We see that the models are able to generate not only CAD programs similar to the geometry conditioning, but also a diverse sample of similar CAD programs. This offers a user options to choose from, should they wish to use one of the generated programs as a starting point. We qualitatively notice that some geometries, particularly complex ones, produce greater variation than others in their generated CAD program samples. We also notice that the mesh generative model produces the most accurate reconstructions but also produces less variation. For more examples of generation, see our project site \url{gencad3d.github.io}.

\subsection{Latent representation of input modalities}

We visualize the latent space of the point cloud + normal model with t-SNE for qualitative analysis. The full latent space with all modalities can be seen in Figure \ref{fig:teaser}. Since the contrastive model groups ``similar' geometries closer together or farther apart, we aim to identify interpretable trends in the embedding space. This helps assess whether the model is simply memorizing the mappings or learning trends. We can see some clustering among prismatic and cylindrical geometries, suggesting that the encoders are indeed learning geometric features.


\begin{table*}[h]
        \centering
        \caption{\textbf{Ablation on Synthetic Augmentation Choice}. (*) denotes our reported SynthBal model.}
   \begin{NiceTabular}{l |c@{\hskip 5pt}c@{\hskip 5pt}c@{\hskip 5pt}c@{\hskip 5pt}c@{\hskip 5pt} |[tikz=dashed] c@{\hskip 5pt}c@{\hskip 5pt}c@{\hskip 5pt}c@{\hskip 5pt}c@{\hskip 5pt}}
    \toprule
        Augmentation  & $\mu_{\text{cmd}} (\uparrow)$ & $\mu_{\text{param}} (\uparrow)$ & $M_{\text{CD}} (\downarrow)$ & $IoU (\uparrow)$ &  $\text{IR} (\downarrow)$ & $\mu_{\text{cmd}}^{\text{SL}} (\uparrow)$ & $\mu_{\text{param}}^{\text{SL}} (\uparrow)$ & $M_{\text{CD}}^{\text{SL}} (\downarrow)$ & $IoU^{\text{SL}} (\uparrow)$& $IR^{\text{SL}} (\downarrow)$\\
    \midrule
        Noisy Replace-Sketch (SynthBal)*  & 99.68 & 98.22 & 0.769 & \textbf{95.1}& 0.845 & 98.57 &93.56  & 2.72 & \textbf{76.5}& 3.36\\ 
        Maximize Real Data  &99.59& 98.05 & 0.764 & 94.5& 0.832 & 98.29 &94.00  & 3.01  &76.1& 4.68\\
        No Validity Check & 99.71  & \textbf{98.53} & \textbf{0.743} & 95.0& 2.372& 98.67 &\textbf{94.25}  & 2.52  & 74.4& 6.31 \\  
        Reduced Pure-Noise   &99.58 & 98.31 & 0.767 & \textbf{95.1}& 0.869 & 98.21 &93.62  & 3.47 & 74.8& \textbf{3.19}\\ 
        No Pure-Noise        &99.56 & 98.08 & 0.786& 94.7& 0.907 & 98.36 & 93.38 & 2.91 & 74.2& 4.29\\ 
        Noiseless   & 99.79 &97.42 &0.868 & 91.4& 0.956 &98.87  & 91.51 & 10.52 & 66.9& 5.13 \\ 
        Noise-Only    &99.20 & 97.49& 0.795 & 93.8& 1.764 & 95.79 & 91.50 & 5.58 & 72.3& 8.62\\ 
        Noisy Replace-Extrude   & 99.60 &98.36 &0.750 &94.7 &\textbf{0.708} &98.40 &94.12 &\textbf{2.46}& 76.2 & 3.36 \\
        Noisy Re-Extrude  &99.14 & 97.57 & 0.809 & 93.9& 1.925 & 95.69 & 91.35 & 5.27  & 72.9& 11.57\\ 
        Noisy Arc-Augment  & 99.33 &97.55&0.821&93.8&2.236&96.13&91.26&6.38 &72.3 &12.48 \\
        Noisy RRE & 99.92 & 98.26 & 0.774 & 93.9& 2.261& 99.51 & 92.61 & 3.37& 70.2& 6.70 \\  
        RRE & \textbf{99.97}  &97.21& 0.893 & 90.5& 2.136 & \textbf{99.84} & 91.36  & 7.02 & 66.6& 4.57 \\  
        \hline
        ContrastCAD+RRE\cite{jung2024contrastcad} &$99.76$ & $98.58$ & $0.731$ & 95.8 & $2.53$ & $98.24$ & $93.02$ & $4.69$& 73.6 &$ 8.49$\\   
    \hline

    \end{NiceTabular}

      \label{tab:aug_ablation}
\end{table*}

\subsection{Ablation Studies} \label{sec:ablation}
\color{white}
.

\color{black}
\textbf{Ablation: Choice of Synthetic Augmentations.} \label{sec:synthetic-aug}
Here, we assess the effects of using different augmentation strategies to create the SynthBal dataset. We demonstrate that while augmentation choice has an important effect on the model's performance, the primary contribution of the SynthBal augmentation strategy comes from its ability to balance datasets across sequence lengths.

We trained and compared each autoencoder without the fine-tuning step.  The augmentation strategies are listed below:
\begin{itemize}
    \item Noisy Replace-Sketch: Our baseline augmentation scheme used for SynthBal. Recall that this refers to randomly applying one of two augmentations: large pure-noise or small noise followed by a sketch-replace. We limit the ratio of real data per sequence length, $r$, to 0.2. All the following ablations are constructed as modifications of this augmentation scheme.
    \item No Validity Check: We skip checking for whether the augmented CAD compiles into a valid B-Rep.
    \item Maximize Real Data: We limit the ratio of real data $r$ to 1.0 instead of 0.2, maximizing the amount of real data present in the synthetic dataset.
    \item Reduced Pure-Noise: instead of applying large pure-noise, we apply small pure-noise.
    \item No Pure-Noise: We do not apply any pure-noise and only augment with small noise + replace-sketch.
    \item Noiseless: We only augment with replace-sketch and do not add noise.
    \item Noise-Only: We only augment by perturbing parameters with small noise.
    \item Noisy Replace-Extrude: Instead of replace-sketch, we use replace-extrude. Similar to replace-sketch, except we replace sketch-extrude pairs instead of only sketches between two CAD programs.
    \item Noisy Re-Extrude: Instead of replace-sketch, we use re-extrude. This randomizes the parameters of extrude commands.
    \item Noisy Arc-Augment: Instead of replace-sketch, we use arc-augment. This randomizes the parameters of arc sketch commands.
    \item Noisy RRE: Instead of replace-sketch, we use RRE, the augmentation described by ContrastCAD \cite{jung2024contrastcad}. This consists of three different augmentations applied in series: replace-extrude, re-extrude, and arc-augment. We skip the validity check: RRE produced significant deviations that resulted in predominantly invalid CAD programs, which made it infeasible to generate a validity-checked dataset.
    \item RRE: We apply RRE only with no noise. We also skip the validity check.

\end{itemize}

We assess performance based on reconstruction accuracy of the autoencoder (Table \ref{tab:aug_ablation}). We see that there is no augmentation strategy that outperformed all others in all metrics. Instead, the reconstruction performance constitutes a Pareto front by which one metric can be improved at the cost of another. For example, the RRE augmentation strategy produced the best command accuracy but the second worst parameter accuracy and chamfer distance. Nevertheless, most of the augmentations performed better than the baselines from the literature. 

The augmentation strategies that incorporate commands between multiple CAD programs (replace-sketch and replace-extrude) tended to perform better in command accuracy than those that did not (Only Noise, Noisy Arc-Augment, Noisy Re-Extrude). We believe that this occurs because swapping sketch/extrude portions introduces the autoencoder to a greater variety of sequences of commands, as opposed to the other methods that only vary parameter values. 

We also noticed that methods that incorporated some noise tended to produce better parameter accuracy than those that did not. For example, we see that the No Pure-Noise model had better parameter accuracy than the Noiseless model, and both the SynthBal and Reduced Pure-Noise models had better parameter accuracy than the No Pure-Noise model. This is expected because increased noise creates more variation in parameter values.

Our RRE strategies, particularly Noisy RRE, produced the best command accuracies in both normalized and unnormalized metrics. These even outperformed ContrastCAD, which originally introduced RRE. 

When we did not validate CAD models, we sacrificed valid reconstructions for more accurate reconstructions. Comparing the SynthBal model to the No Validity Check model, we see that removing validity checks produced better command accuracy, parameter accuracy, and chamfer distance across both normalized and unnormalized metrics, but the invalid ratio significantly worsened (88\% increase in normalized metrics). We hypothesize that one source for this phenomenon is that the DeepCAD dataset contains some invalid CAD programs, which we trained our validity-checked autoencoders to avoid creating.

We also see that populating the synthetic dataset with as many real datapoints as possible did not improve performance. Comparing SynthBal to the Maximize Real Data model, we see that SynthBal, populated with $r=0.2$, produced better sequence-length normalized command accuracy, chamfer distance, IoU, and invalid ratio. We hypothesize that this occurs because our synthetic augmentations produce a greater variety of CAD programs than exist in the original "real" dataset, especially at low sequence lengths where most parts are simple variations of prisms or cylinders. Thus, reducing the proportion of real data provides a more diverse dataset on which to train.

\begin{table*}[h]
    \centering
        \caption{\textbf{Ablation of SynthBal on Contrastive Model}. Performed using the Pointcloud + Normals Modality }
    \begin{tabular}{cc|llll}
    \toprule
    Autoencoder & Contrastive Dataset & $R_B=10$ & $R_B=128$ & $R_B=1024$ & $R_b=2048$ \\
    \midrule
    
    Original  &  Original &  $100.0 $   & $96.76 $    & $83.56 $     & $75.53 $ \\
     SynthBalFT  &  Original &  $99.8$   & $\boldsymbol{96.94}$    & $84.75$     & $77.73$ \\
     SynthBalFT  &  SynthBal & 99.0  & 92.38  & 71.00  & 60.88 \\ 

     *SynthBal-1MFT  &  Original &  $99.4 $   & $96.81 $    & $\textbf{87.25} $     & $\textbf{80.84} $ \\
     SynthBal-1MFT  &  SynthBal & 99.2   & 92.70    & 72.49   & 62.92 \\    

    \bottomrule \\
    
    \end{tabular}
    \label{tab-retrieval-abl}
\end{table*}

\begin{table*}[h]
        \centering
        \caption{\textbf{Ablation of SynthBal on Generative Model}. Performed using the Pointcloud + Normals Modality}
    \begin{NiceTabular}{ccc|lllll}
    \toprule
     Autoencoder & Contrastive Dataset & Generative Dataset  & $\mu_{\text{cmd}}^{\text{SL}} (\uparrow)$ & $\mu_{\text{param}}^{\text{SL}} (\uparrow)$ & $M_{\text{CD}}^{\text{SL}} (\downarrow)$ & $IoU^{\text{SL}} (\uparrow)$& $IR^{\text{SL}} (\downarrow)$ \\ 
    \midrule
    
    Original  & Original & Original  & 48.35  &  43.33 &  59.56&  38.8 &22.56 \\ 
    SynthBalFT  & Original & Original  & 50.99  &  47.31 &  56.65&41.7 & 18.35 \\
    SynthBalFT  & SynthBal & Original  &53.82  &  48.30  &  59.93 & 43.2& \textbf{15.75} \\
    SynthBalFT  & SynthBal & SynthBal  &51.91  &  42.17 & 75.38& 33.2& 22.22 \\
    SynthBalFT  & Original & SynthBal  &48.27  &  37.34 &  96.95& 30.8 & 18.97  \\ 
    SynthBal-1MFT  & Original & Original  & 55.73   & 50.62   & 46.03  &\textbf{44.8}   & 22.09 \\ 
    SynthBal-1MFT  & SynthBal & Original  & \textbf{56.42}  &  \textbf{51.15} &   \textbf{45.80 }& 44.6 & 25.50  \\

    \bottomrule 
    \end{NiceTabular}

      \label{tab:gen_ablation}
\end{table*}

\textbf{Ablation: Using Synthetic Data during Autoencoder, Contrastive, and Generative Training Steps.}
The GenCAD-3D architecture consists of three training steps. For each step, we may choose to train with the SynthBal dataset or the original dataset. We perform ablations, varying the choice of autoencoder, contrastive training dataset, and diffusion training dataset, to assess their effects on performance in retrieval and conditional reconstruction. We train all of these models in the point cloud + normal modality. For a baseline, we use the models trained fully with the Original dataset.

We did not perform ablations using the SynthBal-1M dataset during the contrastive or generative training steps. Although training with the larger dataset may improve performance, the process of compiling 1 million meshes and point clouds would be resource intensive. That would reduce the scalability of our augmentation scheme, which focuses on training a high-quality autoencoder that can be reused to improve downstream tasks.

With respect to retrieval (Table \ref{tab-retrieval-abl}), we see that training the contrastive model with Synthetic autoencoders and the Original dataset produces the best retrieval. However, training the contrastive model with the SynthBal dataset significantly decreased performance, with at least a 51\% drop in retrieval performance from the baseline from 2048. This result comes in contrast to the results of autoencoder training, where we found that an autoencoder trained with synthetic data successfully transferred its performance to "real" data. 

We hypothesize that the contrastive model fails to transfer to real data due to dissimilarity in distribution between the synthetic and real geometric datasets. Model transfer between two domains tends to lose effectiveness as the domains become too "different," even when the modality is kept the same \cite{neyshabur2021transferredtransferlearning}. We believe that the synthetic geometric dataset differs significantly enough in distribution from the real geometric dataset to prevent proper transfer, whereas the synthetic CAD program dataset maintains a similar distribution to the real one.

With respect to conditional reconstruction (Table \ref{tab:gen_ablation}), we see that training schemes using the SynthBal dataset in the contrastive step and the Original dataset in the generative step produced the best performance. Across models trained with the SynthBalFT autoencoder, this scheme produced the best improvement in command accuracy (11\%), parameter accuracy (9\%), IoU (7\%), and invalid ratio (30.2\%) from the baseline. However, this also came with a slight increase in chamfer distance (0.6\%). Across models trained with the SynthBal-1MFT autoencoder, we similarly see that this scheme produces the best improvement in command accuracy (16\%), parameter accuracy (14\%), and chamfer distance (23\%). However, the SynthBal-1MFT model trained with the original dataset in the contrastive and generative steps produced better improvements in IoU (10\%) and invalid ratio (2\%). Across all models, while no model produced the best performance in all metrics, the models trained with the SynthBal-1MFT autoencoder have the best performance in most metrics.

Overall, we see that training with a synthetically balanced dataset, especially in the autoencoder stage, improves performance. We also see that training with a larger dataset, enabled by synthetic data generation, improves performance. Across retrieval and generative reconstruction results, we may extrapolate that using synthetic balanced datasets to train upstream models and training the final "task-performing" model with real data produces the best performance.

\subsection{Translation to GUI-Based CAD Software for User Editing}  

\begin{figure*}[!ht]
    \centering
    \includegraphics[width=0.9\linewidth]{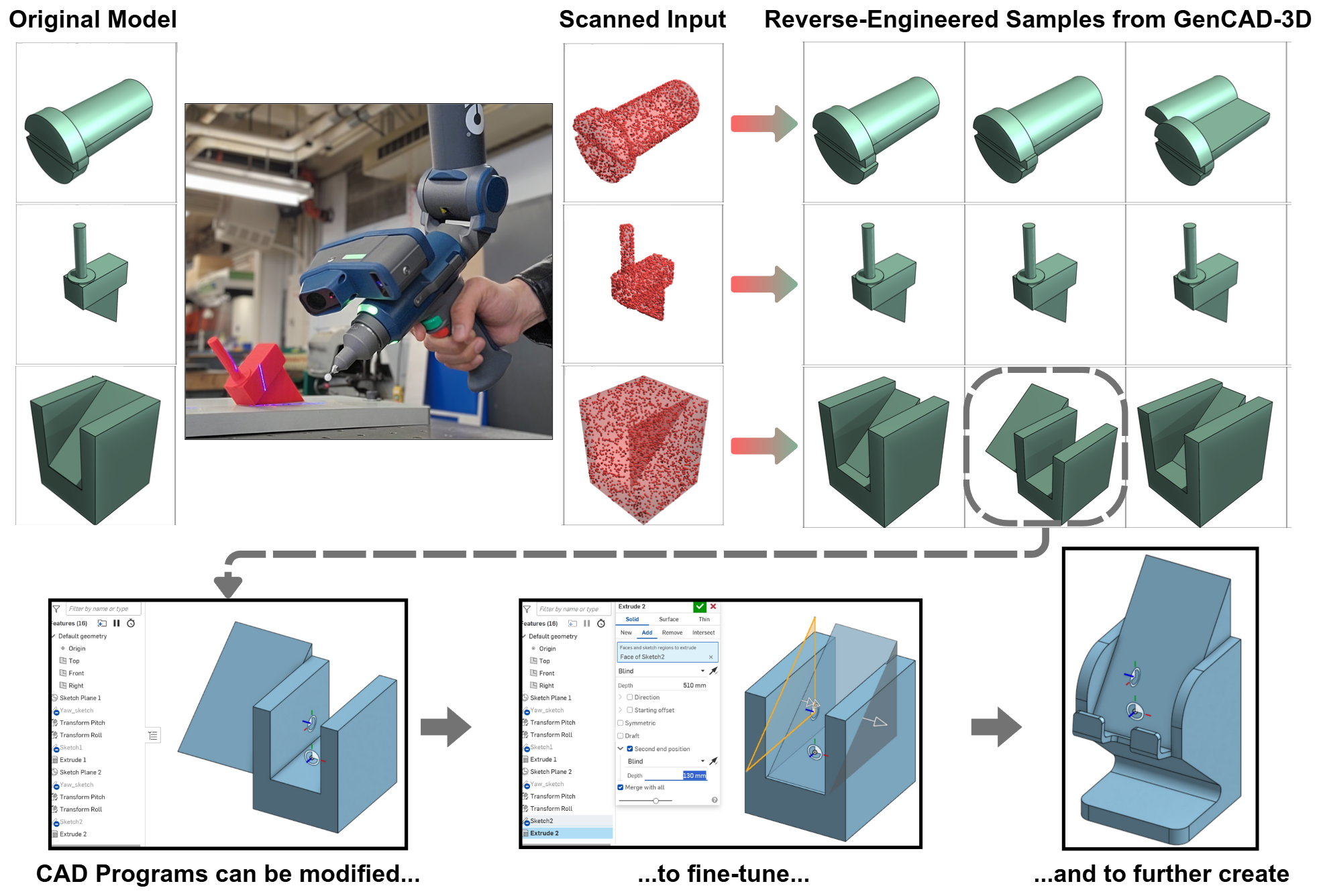}
    \caption{\textbf{Generative Model Shows Reverse-Engineering Capabilities With Real Scanned Point Clouds.} Using our tool, a user can further translate the generated CAD program to a commercial software with a user-friendly GUI so that they can fine-tune the output or use it as a starting point for a larger design.}
    \label{fig:scan2cad}
\end{figure*}

Although generated CAD programs are parametric and low-dimensional, they are still in an encoded format, making them difficult to edit, especially without visual feedback. We created a tool to translate encoded CAD programs to Onshape \cite{onshape}, a free commercial CAD software, to provide a GUI for ease of user editing. We first interpreted the encoded CAD program, which quantized all values, into their corresponding commands and "real" values using tools provided by \cite{wu2021deepcad}. We then used Onshape's web API \cite{onshapeWebApi} and Featurescript language \cite{onshapeFeatureScript} to translate each command into its corresponding definition in Onshape and sequentially send them into an Onshape document. Fig. \ref{fig:scan2cad} shows the CAD program translated into the Onshape environment. We highlight the Feature Tree in the left hand column, which provides users with direct access to any of the parameters in the original CAD program. We present a video showing real-time translation here \url{https://youtu.be/ydStyMHnN-k}.

This tool allows for two important capabilities. First, it allows users to leverage the generative model to automate the bulk of CAD program creation since inaccuracies can now be corrected in a user-friendly editing environment. Second, the translation allows users to generate CAD programs compatible with larger projects maintained on a commercial CAD platform.

\begin{table*}[h]
        \centering
        \caption{\textbf{Scan Reconstruction Evaluations}.}
    \begin{NiceTabular}{cc|lllll}
    \toprule
     Modality & Autoencoder & $\mu_{\text{cmd}}^{\text{SL}} (\uparrow)$ & $\mu_{\text{param}}^{\text{SL}} (\uparrow)$ & $M_{\text{CD}}^{\text{SL}} (\downarrow)$ & $IoU^{\text{SL}}(\uparrow) $& $IR^{\text{SL}} (\downarrow)$ \\ 
    \midrule
    
    Pointcloud  & SB-1MFT (SB)  & 58.54 & 55.54 &  31.0 & 51.3& 15.69  \\ 
    PC+N  & SB-1MFT (SB)  & 34.08 & 23.88 &  173.6 & 27.9& 52.9  \\ 
    Mesh  & SB-1MFT (SB)  & 32.14 & 26.98 &  138.4 & 28.9& 47.1 \\ 
    \bottomrule 
    \end{NiceTabular}

      \label{tab:scan_reconstruction}
\end{table*}

\subsection{Reverse Engineering Point Clouds}

We also demonstrate reverse-engineering of CAD programs from scanned point cloud data. In support of this and to enable future efforts, we provide a test dataset of 51 scanned parts, which were 3D printed from selected models from DeepCAD. We also present a demonstration of our reconstructions applied to those models. 

The physical process of scanning introduces artifacts (noise, occlusions, decreased scan accuracy at certain feature types like edges) that cannot easily be replicated virtually, and so we release this dataset to allow assessments of reverse-engineering methods in more realistic scenarios. To identify the 51 models, we sought a mix of complex and simple geometries that span a variety of command types and geometric features. The scans were translated and scaled to align with the original CAD program.

\begin{figure*}[!ht]
    \centering
    \includegraphics[width=0.8\linewidth]{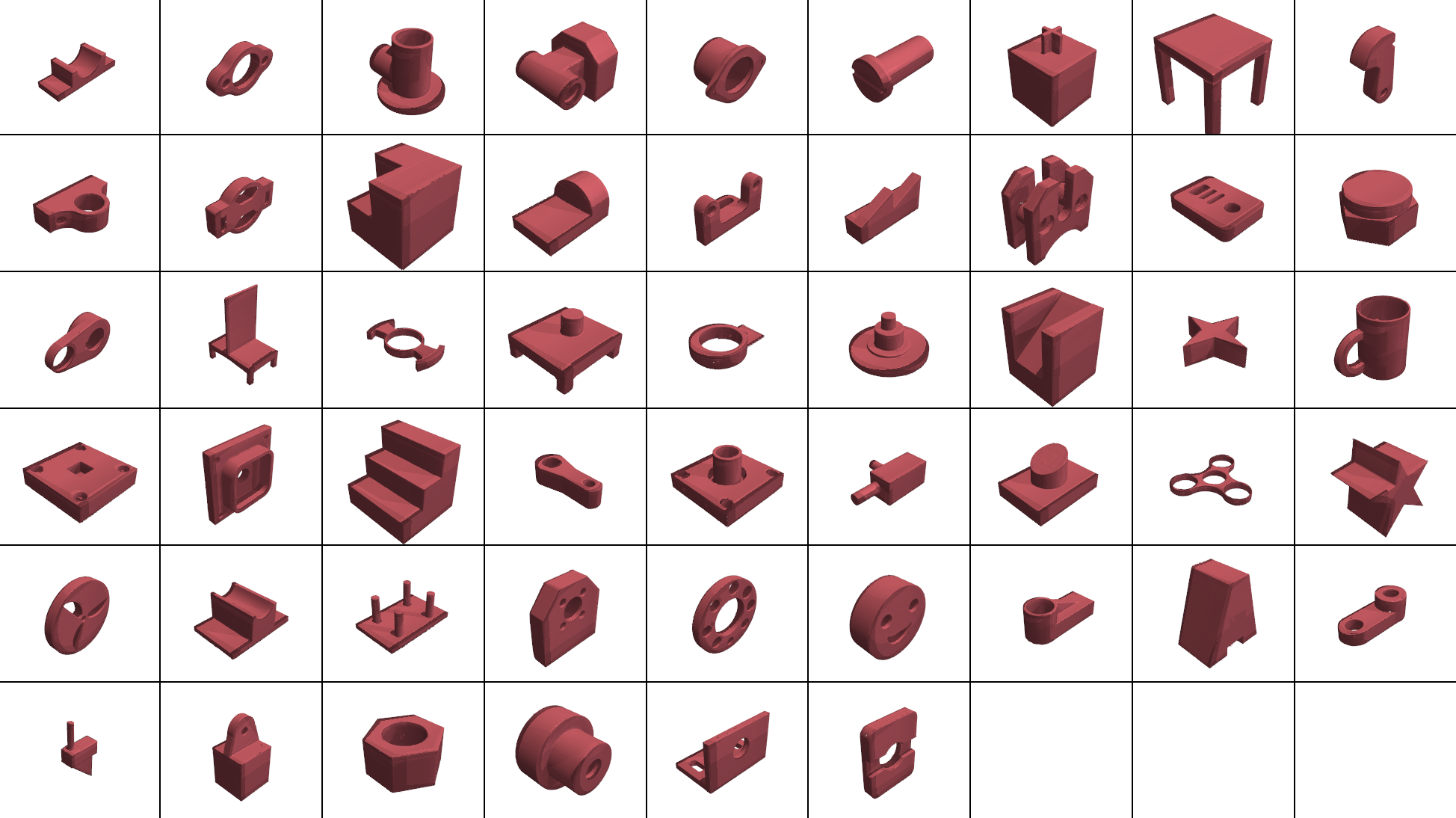}
    \caption{\textbf{Scanned Dataset of 51 3D Printed Parts from DeepCAD}}
    \label{fig:scan-dataset}
\end{figure*}

To enable creation of the scan dataset, we 3D printed the parts on a Bambu X1-Carbon printer, normalizing the maximum bounding box dimension of each part to 60 mm and using a layer height of 0.2 mm. We then scanned all parts on a Faro Arm Quantum S Max scanner \cite{faroArm}. When each part is placed in a specific orientation during scanning, some regions of the part are occluded by the ground. To address this, we scanned each part in 3 orientations, aligned each scan using the PolyWorks Iterative Closest Point (ICP)-based alignment tool \cite{polyworks}, and merged each point cloud together into a single scan. The output of this process is a mesh constructed from the points collected during the scans. To normalize position and scale, we aligned all parts to the original CAD program by again using the Polyworks alignment tool. We finalize by cleaning the scan mesh to remove floating surfaces. The final selection of 51 parts is shown in Figure \ref{fig:scan-dataset}.

We then prepare the scans as inputs into the generative model in a similar way we produced the training dataset. For the pointcloud and pointcloud+normal model, we sample a collection of 2048 points from the scan mesh surface. For the mesh model, while the scan output is in the form of a mesh, the vertex count is high, and so we remesh using a single iteration of isotropic remeshing to reach a vertex count between 2000-4000.

We assess the quality of reverse-engineering by evaluating the generative model on the 51 test set scans. The reconstruction accuracy of each modality is displayed in Table \ref{tab:scan_reconstruction}. We find that the pure-point cloud generation model produced the best reconstructions across all metrics, and the point cloud+normal and mesh models show no clear superiority between each other. We believe the issue lies in the normals: the surface normals of the scan surfaces are likely noisier than those found in the training dataset. We show a selection of parts reverse-engineered from the point cloud model in Figure \ref{fig:scan2cad}. We qualitatively see that the model can produce parts that are similar to the original but that provide variation for a user to choose from.

\section{Implications for Design}

The methods presented in this paper have significant implications for engineering design workflows, particularly in improving efficiency and precision during the design and reverse-engineering phases. By introducing a multimodal latent space capable of translating geometric modalities such as point clouds and meshes into editable CAD programs, designers and engineers can significantly reduce the time required to convert non-parametric geometries into usable parametric CAD models. This facilitates faster iterations, greater flexibility in exploring design alternatives, and ultimately accelerates the overall product development process.

Moreover, SynthBal, a balanced synthetic dataset generation approach, addresses critical dataset imbalance issues, particularly facilitating the design of complex, high-dimensional CAD models. This not only improves neural model performance but also ensures a fairer and more robust evaluation of CAD-generation methods across varying levels of complexity. These advancements enhance the reliability and applicability of generative models in practical engineering design.

\section{Limitations, Future Work, and Conclusion}
In conclusion, our paper introduces GenCAD-3D, a multimodal latent space framework that successfully generates parametric CAD sequences from geometric inputs. Experimental results show significant improvements in retrieval accuracy, CAD reconstruction, and generation performance, particularly for higher sequence complexities. These results validate our contributions, including modality-specific encoders, the SynthBal dataset, and the sequence-length normalization metric. These advancements represent a significant step toward automating precise CAD generation, paving the way for more efficient engineering design processes.

While our experiments demonstrate significant performance improvements, certain limitations remain. Although SynthBal enhances model performance on the DeepCAD dataset, synthetic data may not fully capture the complexities inherent in industrial designs, potentially limiting generalization. Likewise, the current DeepCAD dataset only contains Sketch-and-Extrude CAD programs, a small subset of engineering geometries. To realize practical CAD generation and reconstruction, it is important to collect and experiment on datasets with larger CAD vocabularies, and perhaps develop architectures better suited to handling special commands such as those that require references to compiled B-Rep geometries.

Future work will focus on refining synthetic data augmentation techniques to better emulate complex real-world geometries, thereby improving model generalizability. Additionally, refining latent embedding strategies to better preserve essential geometric details could significantly improve precision in practical applications. Exploring additional modalities, such as engineering drawings, could further enhance the practical utility of the proposed generative methods.







\bibliographystyle{asmejour}   

\bibliography{_main_arxiv} 



\end{document}